\documentclass[lettersize,journal]{IEEEtran}
\usepackage{amsmath,amssymb,amsfonts,amsthm}
\usepackage{algorithmic}
\usepackage{array}
\usepackage[caption=false,font=footnotesize,labelfont=rm,textfont=rm]{subfig}
\usepackage{cite}
\usepackage{booktabs}
\usepackage{multirow}
\usepackage{textcomp}
\usepackage{stfloats}
\usepackage{url}
\usepackage{verbatim}
\usepackage{graphicx}
\usepackage{xcolor}

\def\BibTeX{{\rm B\kern-.05em{\sc i\kern-.025em b}\kern-.08em
    T\kern-.1667em\lower.7ex\hbox{E}\kern-.125emX}}
\usepackage{balance}

\DeclareMathOperator{\sinc}{sinc}

\theoremstyle{plain} 
\newtheorem{theorem}{Theorem} 
\newtheorem{proposition}[theorem]{Proposition} 
\newtheorem{corollary}[theorem]{Corollary} 

\theoremstyle{definition} 

\theoremstyle{remark} 

\begin{document}
\title{Beamspace Dimensionality Reduction for  Massive MU-MIMO: Geometric Insights and Information-Theoretic Limits}
\author{\IEEEauthorblockN{
Canan Cebeci, Oveys Delafrooz Noroozi and Upamanyu Madhow}\\
\IEEEauthorblockA{
ECE Dept., University of California, Santa Barbara, CA, U.S.A.}\\
\{ccebeci, oveys, madhow\}@ucsb.edu}

\maketitle

\begin{abstract}
Beamspace dimensionality reduction, a classical tool in array processing, has been shown in recent work to significantly reduce computational complexity and training overhead for adaptive reception in massive multiuser (MU) MIMO.  For sparse multipath propagation and uniformly spaced antenna arrays, beamspace transformation, or application of a spatial FFT, concentrates the energy of each user into a small number of spatial frequency bins.  Empirical evaluations demonstrate the efficacy of linear Minimum Mean Squared Error (LMMSE) detection performed in parallel using a beamspace window of small, fixed size for each user, even as the number of antennas and users scale up, while being robust to moderate variations in the relative powers of the users.  In this paper, we develop a fundamental geometric understanding of this ``unreasonable effectiveness'' in a regime in which zero-forcing solutions do not exist. For simplified channel models, we show that, when we enforce a suitable separation in spatial frequency between users, the interference power falling into a desired user's beamspace window of size $W$ concentrates into a number of dominant eigenmodes smaller than $W$, with the desired user having relatively small projection onto these modes. Thus, linear suppression of dominant interference modes can be accomplished with small noise enhancement.  We show that similar observations apply for MIMO-OFDM over wideband multipath channels synthesized from measured 28 GHz data. We propose, and evaluate via information-theoretic benchmarks, per-subcarrier reduced dimension beamspace LMMSE in this setting.

\end{abstract}


\section{Introduction} \label{sec:introduction}
The ability to scale massive MU-MIMO is critical for unlocking the potential of higher carrier frequencies in next-generation wireless systems. More array elements can be packed into the same form factor at smaller wavelengths, and this increase in spatial degrees of freedom can be used to support a larger number of users simultaneously in a given time-frequency resource block.  This promises significant increases in wireless network capacity, especially when coupled with the larger bandwidths typically available as we increase the carrier frequency.  While there are many technical and commercial hurdles to realizing this promise, we focus here on the fundamental challenge of scaling signal processing architectures and algorithms as the number of antennas $N$, number of simultaneous users $K$, and bandwidth $B$ get large.  Even conventional least squares adaptation for the ``simple'' adaptive linear MMSE (LMMSE) receiver for an $N$-element array in a typical narrowband setting is challenging, requiring formation and inversion of an $N \times N$ sample covariance matrix and $O(N)$ training symbols per user.  Fortunately, in many settings of interest, especially at higher carrier frequencies, propagation is dominated by a small number of paths, which opens up the possibility for drastic savings in complexity and overhead via dimension reduction in ``beamspace.''  The goal of the present paper is to develop fundamental geometric insights, design guidelines, and benchmarks for such strategies.

\begin{figure}[h]
\setlength{\tabcolsep}{0pt} 
  \begin{tabular}{cc}
    \includegraphics[width=1\linewidth]{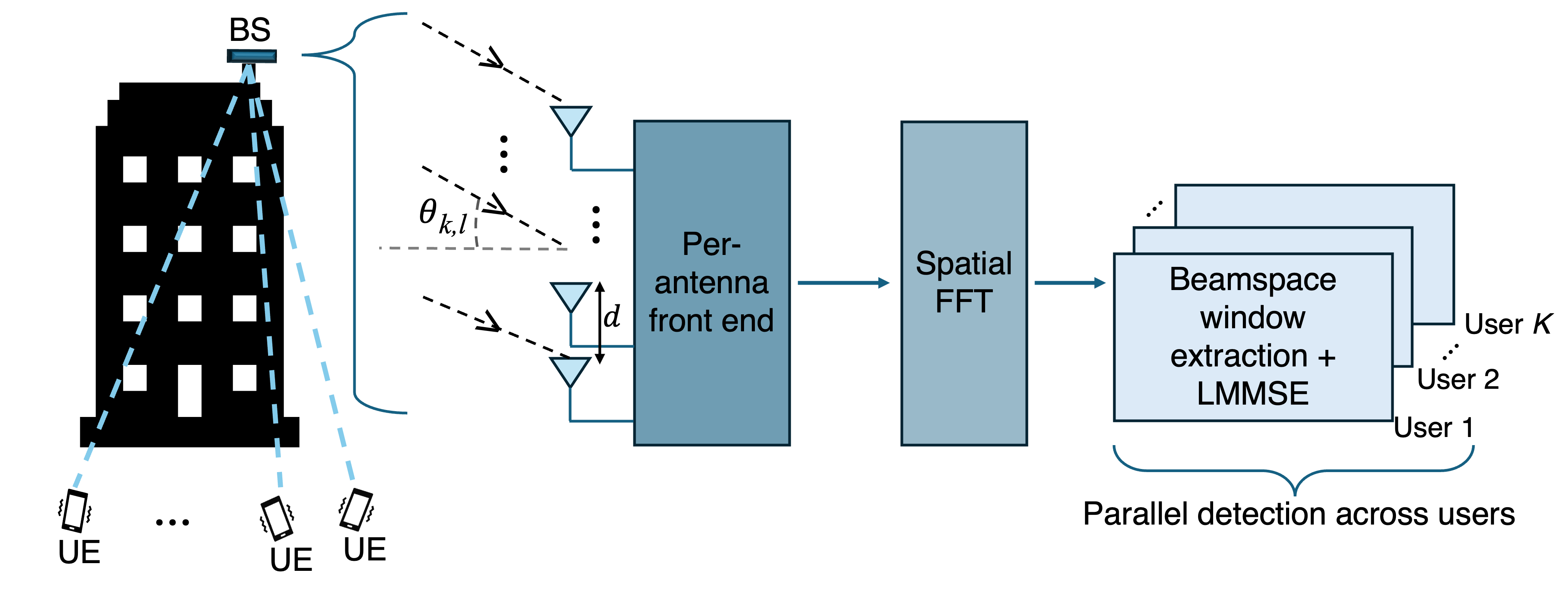}  
  \end{tabular}
  \caption{Uplink massive MU-MIMO system model with beamspace dimensionality reduction. We assume far field for each path reaching at the antenna array.}
\label{systemModelBasic}
\end{figure}

For uniformly spaced antenna arrays, the response to a narrowband signal arriving from any given direction is a complex exponential with spatial frequency depending on the direction of arrival. ``Beamspace transformation'' via a spatial DFT concentrates the energy of any such signal into a small number of spatial frequency bins.  Dimension reduction using this well known concentration property has been utilized for computationally efficient array signal processing for decades.  Empirical evaluations in recent work \cite{localLMMSEAbdelghany,Abdelghany20_scalable_nonlinear,tiledbeamspace,cebeci2024} indicate the potential of this approach for all-digital massive MU-MIMO over sparse multipath channels.   In particular, even as the number of antennas $N$ and number of simultaneous users $K$ scale up proportionally, simulations show that a narrowband system employing parallel adaptive LMMSE reception using a \emph{fixed} beamspace window $W$ (i.e., not scaling with $N$) for each user attains, in certain regimes, performance comparable to conventional ``antenna space'' LMMSE, while drastically reducing the computational complexity and overhead for training. It also exhibits a degree of resilience to power disparities among users. Reduced dimension beamspace LMMSE implemented in parallel for $K$ users in this fashion (see Fig. \ref{systemModelBasic}) incurs $O(K W^3)$ complexity (versus $O(N^3)$ for antenna space) and requires $O(W)$ training symbols (versus $O(N)$ for antenna space) per user. 

Even for simple line of sight (LoS) channels, the system in Fig. \ref{systemModelBasic} is inherently interference-limited.  The spatial frequency for any given interferer falls, with probability one, {\it off the DFT grid,} hence each interferer has nontrivial (albeit attenuated) contributions to the designated beamspace window for a desired user.  In our scaling regime of interest, therefore, the available signal space dimension $W$ is much smaller than the number of interferers $K-1$.  Why, then, does linear interference suppression work so well in this regime?

\subsection{Contributions}

In this paper, we develop a geometric understanding and explanation for the ``unreasonable effectiveness'' of beamspace dimension-reduced LMMSE, and identify system design prescriptions for desired regimes of operation.  We also propose and evaluate, using information-theoretic metrics computed based on measured 28 GHz data, a natural extension of the narrowband dimension-reduced architecture in Fig. \ref{systemModelBasic} to per-subcarrier dimension-reduced LMMSE for wideband MIMO-OFDM.  Our key results are summarized as follows:\\
(1) {\it Desired signal concentration:} We establish a lower bound independent of $N$ on the fraction of energy from a single path captured by a beamspace window of size $W$. The bound implies, for example, that regardless of array size, more than 90\% of the energy is captured by a beamspace window of size $W=4$ centered around the spatial frequency of the path. \\
(2) {\it Interference concentration:} For a simplified LoS channel model, we show that, when a minimum spatial frequency separation is enforced between users, the bulk of the interference energy in a desired user's beamspace window of size $W$ concentrates onto a number of dominant eigenmodes smaller than $W$.  Moreover, the desired user's response has relatively small projection onto these dominant eigenmodes. Thus, the attenuation of dominant interference modes by the LMMSE receiver leads to relatively small loss in the desired user's energy, explaining why it works well in an interference-limited regime.\\
(3) {\it Lower bound on expected SINR and design consequences:} We establish a lower bound on expected SINR and show that, under the preceding simplified model, it depends only on the ratio of the desired user's power to the total interference power, thus capturing the effect of both the number of users and power disparities among users. This yields design guidelines for resource allocation and power control: by enforcing a minimum spatial frequency separation between users which scales as $1/N$, the number of simultaneous users can scale linearly with the number of antennas while using a small, fixed beamspace window to demodulate each user.\\
(4) {\it Beamspace dimension-reduced wideband MIMO-OFDM:} We propose per-subcarrier beamspace dimension-reduced LMMSE for wideband MIMO-OFDM over sparse multipath channels.  For channel models derived from measured 28 GHz data, we empirically observe signal and interference concentration across subcarriers and users when enforcing minimal spatial frequency separation between the dominant paths of the users.\\
(5) {\it Information-theoretic benchmarks based on measured data:} We establish information-theoretic benchmarks for the preceding wideband system, comparing the spectral efficiency of the proposed dimension-reduced linear reception against that of an unrealizable, unconstrained system. \\
We conclude with a discussion of the design implications of the preceding fundamental insights.

\noindent \emph{Notation:} Throughout the paper, bold lowercase symbols represent column vectors, while bold uppercase symbols denote matrices. 
The operators $(\cdot)^{\top}$ and $(\cdot)^{H}$ stand for transpose and Hermitian transpose, respectively. 
For functions $f(N)$ and $g(N)$, we write $f(N) = \Theta(g(N))$ if,
for large $N$, $f(N)$ is bounded above and below by constant multiples of $g(N)$.
The discrete Fourier transform (DFT) operator is the unitary matrix $\mathbf{F}_{N_{\mathrm{FFT}}}$, defined by
\begin{equation}\nonumber
\big[\mathbf{F}_{N_{\mathrm{FFT}}}\big]_{m,n} = \frac{e^{-j \frac{2\pi}{N_{\mathrm{FFT}}} (m-1)(n-1)}}{\sqrt{N_{\mathrm{FFT}}}} , \quad m,n=1,\ldots,N_{\mathrm{FFT}}.
\end{equation}

\subsection{Related Work}
The concept of beamspace dimension reduction emerged in the 1970s to tackle the complexity of large adaptive arrays: \cite{reed1974adaptivearrays} demonstrated a partial adaptivity method in which adaptive weighting on a reduced set of beams yields rapid convergence in large arrays, and \cite{chapman1976partial} formalized partial adaptivity for large arrays, showing that near-optimal interference suppression could be achieved by adapting only a subset of all available degrees of freedom. Beamspace dimension reduction for super-resolution in direction of arrival (DOA) estimation was proposed in the 1990s, including for multiple signal classification (MUSIC) in \cite{MUSICwithbeamspace}, and ESPRIT (Estimation of Signal Parameters via Rotational Invariance Techniques) in \cite{ESPIRITtwithbeamspace}.  Beamspace dimension reduction for Space-Time Adaptive Processing (STAP) for radar was discussed in \cite{Ward1994STAP}, where it was demonstrated that using only a few beams can yield near-optimal clutter suppression with far fewer adaptive degrees of freedom. Recent work applies it for scaling to \emph{wideband} all-digital massive MIMO radar \cite{delafrooz2025}. 

While we focus here on all-digital massive MIMO, beamspace processing has been applied to hybrid analog-digital beamforming architectures for mmWave and massive MIMO over the past two decades. Using an analog discrete lens array (DLA) or Butler matrix to transform the $N$-element signals into $N$ orthogonal beams, then selecting or processing only a subset of those beams is the common characteristics of these studies.
A Continuous Aperture Phased MIMO (CAP-MIMO) architecture, where a high resolution lens forms orthogonal beams such that the energy of each user is concentrated mainly in a few beam outputs, was proposed in \cite{CapMIMO}.  
Beam selection algorithms for such lens-based arrays, optimizing which beams to activate to maximize rate for multi-user systems, were investigated in \cite{BradyAnalog,ShenAnalog}.

The architecture in Fig. \ref{systemModelBasic} was first proposed in 2019 \cite{localLMMSEAbdelghany}, where it was termed ``local MMSE'' detection. The same work also provided a plausibility argument for linear scaling of number of users $K$ with $N$ while maintaining a fixed beamspace window of size $W$ per user.  It was shown that, for users uniformly distributed in spatial frequency, the probability of ``outage'', defined as more than $W$ interferers falling into a desired user's beamspace window, is bounded by an expression depending only on $W$ and the load factor $K/N$.  Estimates of probability of outage, supported by simulations, were used to argue that small load factors (e.g., $K/N = 1/8, 1/4$) could be supported.  Unlike the ad hoc definition of outage in \cite{localLMMSEAbdelghany}, we provide an analytical framework that provides explicit design prescriptions based on pessimistic estimates of the SINR attained by LMMSE reception for a given beamspace window.  While we also model user spatial frequencies as uniformly distributed, we additionally assume that the resource allocation layer enforces a minimum spatial frequency separation between users. This enables supporting larger load factors for small window sizes (e.g., $K/N \approx 1/2$ for $W=5$).

Dimension-reduced beamspace techniques have been applied for complexity reduction in nonlinear MU-MIMO detection \cite{Abdelghany20_scalable_nonlinear} and linear precoding \cite{Abdelghany19_beamspace_precoding}. Dimension reduction also leads to reduced overhead: adaptive linear multiuser detection in reduced dimension beamspace outperforms full-dimension ``antenna space'' adaptation when training overhead is limited, as shown by simulations of the tiled beamspace receiver in \cite{tiledbeamspace}, and by the information-theoretic benchmarks in \cite{cebeci2024}. Energy concentration in beamspace also enables scaling of explicit spatial channel estimation \cite{mirfarshbafan2020}. A hardware realization combining adaptive resolution analog-to-digital conversion with beamspace processing and channel estimation was presented in \cite{castaneda2021}. A layered belief propagation detector in the beam domain, exploiting the fact that the post-FFT channel matrix is sparse to reduce complexity of message passing, is proposed in \cite{layeredbeliefpropbeamspace}.

\section{System Model} \label{sec:system_model}
\subsection{General Multi-user Multipath Model}
We consider a massive MU-MIMO uplink scenario in which a base station (BS) is equipped with an $N$-element uniform linear array (ULA) with inter-element spacing $d = \lambda_c/2$ to serve $K$ single-antenna users, as illustrated in Fig. \ref{systemModelBasic}, where $\lambda_c$ denotes the wavelength corresponding to the carrier frequency $f_c$.

The channel of the $l$-th propagation path associated with the $k$-th user is characterized by its complex gain $\alpha_{k,l}$, path delay $\tau_{k,l}$, and angle of arrival (AoA) $\theta_{k,l}$. The reference spatial frequency of this path, defined with respect to the carrier frequency, is given by
\begin{equation}\nonumber
\label{eq:RefOmega}
\Omega^{\mathrm{ref}}_{k,l} = \pi \sin\left(\theta_{k,l}\right),
\end{equation}
and the spatial frequency can be expressed as a linear function of the temporal frequency $f \in [-\frac{B}{2}, \frac{B}{2}]$, where $B$ denotes the total signal bandwidth, as
\begin{equation}
\label{eq:OmegaWRTFreq}
\Omega_{k,l}(f) = \Omega^{\mathrm{ref}}_{k,l}\!\left(1 + \frac{f}{f_c}\right).
\end{equation}

The array steering vector corresponding to a given spatial frequency is defined as
\begin{equation}
\label{eq:SteeringVector}
\mathbf{a}(\Omega) = \big[1\;\; e^{j\Omega}\;\; e^{j2\Omega}\; \cdots\; e^{j(N-1)\Omega}\big]^{\top}.
\end{equation}
Accordingly, the frequency-domain channel of a single path can be written as
\begin{equation}
\label{eq:ChannelFreq}
\mathbf{h}_{k,l}(f) = \alpha_{k,l}\, \mathbf{a}\!\left(\Omega_{k,l}(f)\right) e^{-j2\pi (f_c + f)\tau_{k,l}},
\end{equation}
and the superposition of all $L_k$ paths yields the overall frequency-domain channel of the $k$-th user:
\begin{equation}
\label{eq:ChannelFreqSum}
\mathbf{h}_{k}(f) = \sum_{l=1}^{L_k} \alpha_{k,l}\, \mathbf{a}\!\left(\Omega_{k,l}(f)\right) e^{-j2\pi (f_c + f)\tau_{k,l}}.
\end{equation}
By substituting $f = 0$, the corresponding narrowband channel reduces to
\begin{equation}
\label{eq:ChannelFreqNB}
\mathbf{h}_k = \mathbf{h}_k(0) = \sum_{l=1}^{L_k} \alpha_{k,l}\, \mathbf{a}\!\left(\Omega^{\mathrm{ref}}_{k,l}\right) e^{-j2\pi f_c \tau_{k,l}}.
\end{equation}

\begin{figure}[h]
\setlength{\tabcolsep}{0pt} 
  \centering
    \includegraphics[width=0.8\linewidth]{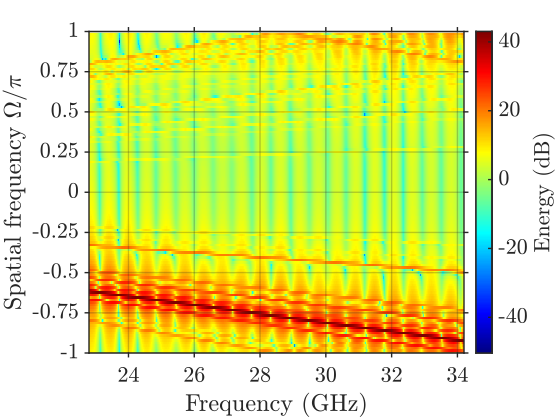}  
  \caption{Energy in the spatial frequency-frequency grid for a single user with 48 paths including the LOS, the BS has $N=128$ antenna elements. The center frequency is 28.5 GHz and the fractional bandwidth is 20$\%$. The channel vector at each frequency is constructed using path-loss, delay, and AoA parameters drawn from the 28.5 GHz channel dataset \cite{charbonnier2020calibration}.}
\label{sparsityWRTfrequency}
\end{figure}

Beamspace dimension reduction is particularly effective for sparse multipath channels with a dominant (often LoS) path along with weaker paths.
Fig. \ref{sparsityWRTfrequency} illustrates such sparsity for measured mmWave channels: most of the received power is concentrated in a narrow cluster of bins centered on the dominant path for each user, which motivates retaining a small window of bins around the most significant bin.  

\subsection{Simplified System Model for the Analysis}

Starting from the channel model in (\ref{eq:ChannelFreqNB}), we note that different paths for desired and interfering users have different delays, so over a given symbol interval, they are modulated by different symbols.
Therefore, we abstract the problem to a simplified setting consisting of one desired user represented by a single propagation path, together with multiple interfering paths, each modulated by mutually uncorrelated symbols:
\begin{equation}
\label{simplifiedmodel}
\mathbf y = \alpha_1 b_1 \mathbf v_1 \;+\; \sum_{k=2}^K \alpha_kb_k \mathbf v_k \;+\; \mathbf w,
\end{equation}
where 
$ \mathbf v_k = \mathbf a(\Omega_k)$, 
$\mathbf w \sim \mathcal{CN}(0, 2\sigma^2 \mathbf I_N)$ is the noise vector, $b_k$ are complex independent symbols with unit energy, $\alpha_k$ are path gains, and  
$\mathbf a(\Omega_k)$ is the array steering vector corresponding to AoA $\theta_k$.

Under any linear transformation $\mathbf z = \mathbf T \mathbf y$, the model becomes
\begin{equation} \label{simplifiedmodel_beamspace}
\mathbf z = \alpha_1b_1 \mathbf u_1 \;+\; \sum_{k=2}^K \alpha_kb_k \mathbf u_k \;+\; \mathbf n,
\end{equation}
where $ \mathbf u_k = \mathbf T \mathbf v_k, 
\mathbf n \sim \mathcal{CN}(0, \mathbf C_n),
\mathbf C_n = 2\sigma^2 \mathbf T \mathbf T^H.
$

The interference–plus–noise correlation matrix is then
\begin{equation}\label{interferenceplusnoise}
\mathbf R_{I+N} = \sum_{k=2}^K P_k \,\mathbf u_k \mathbf u_k^H + \mathbf C_n,
\end{equation}
with $P_k$ denoting the $k$th user's antenna-space receive power before any beamforming gain.  
The ideal LMMSE correlator for the desired user is proportional to the whitened matched filter (\!\!\cite{madhow1994})
\begin{equation}
\mathbf c = \mathbf R_{I+N}^{-1} \mathbf u_1,
\end{equation}
which achieves the best possible SINR among linear receivers:
\begin{equation} \label{sinr}
\mathrm{SINR} = P_1\mathbf u_1^H \mathbf R_{I+N}^{-1} \mathbf u_1. 
\end{equation}

We will be interested in transformations $\mathbf T$ corresponding to beamspace dimension reduction: a spatial FFT followed by per-user windowing.  Our goal is to scale the number of users $K$ linearly with the number of antennas $N$ while keeping the beamspace window dimension $W$ fixed, which ensures that the per-user receiver complexity does not grow with array size.

\section{Signal and Interference Concentration} \label{sec:concentration}
In this section, we develop geometric insights into signal and interference concentration in beamspace. We show that a relatively small beamspace window whose size does not scale with $N$ suffices to capture a substantial fraction of the signal energy from any given ray, and that, provided a minimum spatial frequency separation is imposed between users, the interference concentrates into a small number of modes which can be attenuated without excessive loss of desired signal energy.

\subsection{Signal Concentration} \label{sec:signal_concentration}
For the simplified model (\ref{simplifiedmodel}), consider a desired signal in antenna space corresponding to a single path at spatial frequency $\Omega \in [- \pi, \pi]$: $\mathbf v = \mathbf a(\Omega)$. For our default scenario of beamspace transformation without zeropadding, we take an $N$-point FFT, where the $N$ bins are numbered from $-N/2$ to $N/2-1$.  Any spatial frequency can be written as  $\Omega = 2 \pi (n_0 \pm \delta )/N$, where $n_0$ is the nearest grid point and $\delta \in [0, 0.5]$. Since the DFT is unitary, total energy is preserved from antenna space to beamspace. 
Normalizing the DFT coefficients so that their energy sums to one, the $n$th DFT coefficient is the normalized inner product of the complex exponential $\mathbf a(\Omega)$ against the $n$th element of the DFT basis, given by:
\begin{equation} \label{normalized_DFT}
y_n = \frac{1}{N} \mathbf a^H (\frac{2\pi n}{N}) \mathbf a (\Omega) = D_N \left( \Omega - \frac{2\pi n}{N} \right)
\end{equation}
where $D_N ( \cdot )$ denotes the Dirichlet kernel
\begin{equation} \label{dirichlet_kernel}
D_N ( \omega ) = \frac{1}{N} \sum_{n=0}^{N-1} e^{j \omega n} = e^{j (N-1) \omega/2} \frac{\sin (N\omega/2)}{N \sin(\omega/2)}
\end{equation}
Here $D_N (0) = 1$ is defined as usual by taking the limit as $\omega \rightarrow 0$.  Note that $D_N (\omega ) = 0$ if and only if 
$\omega = 2 \pi k/N$ for some integer $k \neq 0$ (modulo $N$).

For on-grid $\Omega$ (i.e., $\delta = 0$), the entire energy is captured in single bin, $n_0$.
For off-grid $\Omega$ ($0 < \delta < 0.5$),  the energy is spread over all $N$ bins: the Dirichlet kernel evaluated at $\Omega - \frac{2\pi n}{N} = \frac{2\pi (n_0-n \pm \delta)}{N}$ in (\ref{normalized_DFT}) is nonzero for every $n$, and we choose a window around the highest power bin, as follows.

\noindent
{\it Choice of beamspace window $\mathcal W^\star$:}
We define the window of bins for even and odd $W$ separately:\\
$\bullet$ For odd $W = 2 K+1$, the beamspace window consists of bins with indices $W^\star=\{ n_0 - K,...,n_0,...,n_0+K \}$, where the indices are computed modulo $N$.\\
$\bullet$ For even $W = 2K$, the bin indices are 
$W^\star=\{ n_0 - K+1,...,n_0,...,n_0+K \}$ 
if $\Omega = 2 \pi (n_0 + \delta )/N$, and 
$W^\star=\{ n_0 - K,...,n_0,...,n_0+K-1 \}$
if $\Omega = 2 \pi (n_0 - \delta )/N$, with all indices again interpreted modulo $N$.\\
We denote by $\mathbf W_{\mathbf{k}}$ the $W \times N_{\mathrm{FFT}}$ matrix that selects the beamspace bins corresponding to user $k$.

Let $E_{W,N} (\delta )$ denote the fraction of energy captured in a window as defined above for an $N$-element array.  The following theorem, proved in Appendix \ref{app:proof_energy_capture}, provides a lower bound on the energy capture which depends on $W$ but not on $N$.

\begin{theorem}[Lower bound on energy capture with optimal $W$-bin contiguous window placement]
\label{thm:stdW}
For $N \geq W \geq 1$ and any fractional offset $\delta \in [0,0.5]$, the maximum energy captured by an optimally placed contiguous $W$-bin window satisfies
\begin{equation*}
E_{W,N} (\delta) \;\geq\; \sum_{n\in\mathcal W^\star} \sinc^2\bigl(n - (n_0 \pm \delta)\bigr),
\end{equation*}
where $\mathcal W^\star$ is the set of $W$ DFT indices defined above.
\end{theorem}

The minimum of the right-hand side over $\delta \in [0,0.5]$ can be calculated explicitly and is attained at $\delta = 0.5$.

\begin{corollary}[Energy capture with four-bin window]
At least $90\%$ of the desired-user energy is guaranteed to lie within an optimally placed window of size $W=4$.
\end{corollary}
To see this, we numerically evaluate the lower bound in Theorem~\ref{thm:stdW}, which attains its minimum at $\delta = 0.5$, obtaining
\begin{equation}
\sum_{n\in\mathcal W^\star} \sinc^2\bigl(n - (n_0 \pm 0.5)\bigr)
\;=\;
\frac{80}{9\pi^2}
\;\approx\; 0.9006.  
\end{equation}

\subsubsection{Signal concentration with zeropadded DFTs} 

Intuitively, we expect better signal concentration due to the finer DFT grid created by zeropadding, a routine practice in many DFT applications. However, we shall see that this intuition can lead us astray in our present context of dimension-reduced beamspace LMMSE.
Recall that the SINR attained by the LMMSE solution, given by (\ref{sinr}), depends on 
\begin{equation}
\mathbf u_1^H \left( \mathbf R_I + \mathbf R_N  \right)^{-1} \mathbf u_1
\end{equation}
where $\mathbf R_I$ is the interference covariance and $ \mathbf R_N$ is the noise covariance (colored under zeropadding).  While we must account for the coloring when we quantify signal capture in a noise-limited regime, we can ignore $\mathbf R_N$ (and hence its coloring) in interference-limited regimes in which performance is dominated by the geometric relationship between $\mathbf u_1$ and $\mathbf R_I$. We consider these two regimes separately, focusing on zeropadding by a factor of two.

\noindent
{\bf Zeropadding in an interference-limited regime:} We do not need to account for noise coloring in this regime, hence a simple extension of 
Theorem \ref{thm:stdW} can be applied.
The following proposition (proved in Appendix \ref{app:proof_energy_capture}) provides a corresponding lower bound for $2\times$ zeropadding.
\begin{proposition}[Energy capture with $2\times$ zeropadding]
For any $2N \ge W \ge 1$ and any $\delta \in [0,0.5]$,
\begin{equation*}
E^{(\mathrm{zp2})}_{W,N}(\delta)
\;\ge\;
\frac{1}{2}
\sum_{n\in\mathcal W^\star}
\sinc^2\!\left(\frac{n}{2} - (n_0 \pm \delta)\right)  
\end{equation*}
where the factor $\tfrac{1}{2}$ arises from the normalized Dirichlet-kernel representation of the $2N$-point DFT.
\end{proposition}
In order to fully understand whether or not zeropadding helps or hurts in this regime, however, we must also investigate the structure of the interference, as we do in Section \ref{sec:interference concentration}.

\begin{figure}[h]
\centering
\subfloat[]{%
\includegraphics[width=0.5\linewidth]{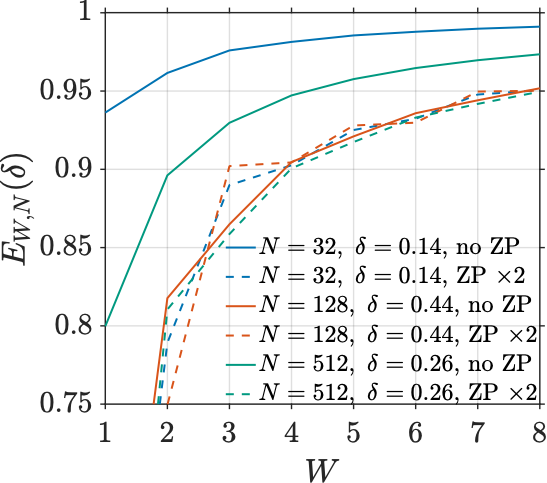}%
}\hspace{0.em}%
\subfloat[]{%
  \includegraphics[width=0.5\linewidth]{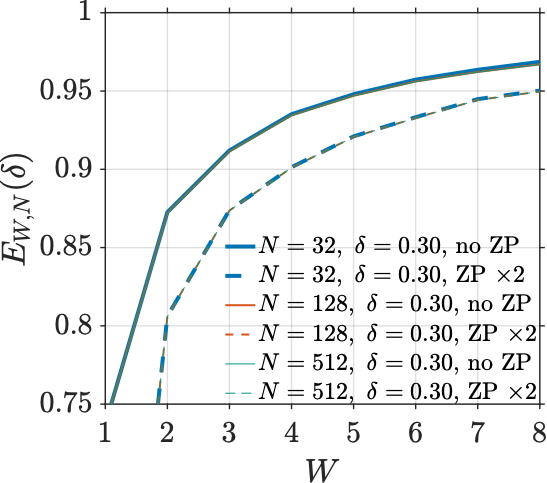}%
}\\[-0.15em]
\subfloat[]{%
\includegraphics[width=0.6\linewidth]{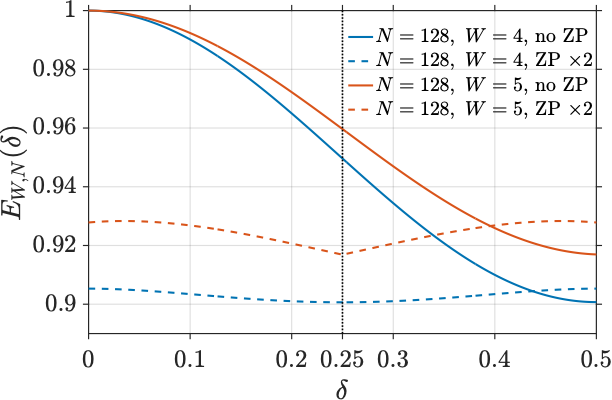}%
}
\caption{Desired energy ratio with and without zeropadding in an interference-limited regime.
(a) Fixed arrival angle ($\theta^\circ=15$), so the spatial-frequency offset to the DFT grid varies with $N$.
(b) Fractional off-grid offset held constant ($\delta=0.3$) across $N$.
(c) $E_{W,N}(\delta)$ and $E^{(\mathrm{zp2})}_{W,N}(\delta)$, for $N=128$.}
\label{plt:desiredenergycapturewithoutnoise}
\end{figure}

Fig. \ref{plt:desiredenergycapturewithoutnoise} reports empirical energy-capture curves $E_{W,N}(\delta)$ for a $W$-bin window across three array sizes. The capture increases rapidly with $W$ and agrees with the theorem’s prediction. 
With a fixed AoA (Fig. \ref{plt:desiredenergycapturewithoutnoise}a), the fractional off-grid offset $\delta$ varies with $N$; when $\delta$ is held fixed across arrays (Fig. \ref{plt:desiredenergycapturewithoutnoise}b), the curves align, showing that for contiguous windows $E_{W,N}(\delta)$ is governed primarily by $(W,\delta)$. 
Fig. \ref{plt:desiredenergycapturewithoutnoise}c compares the energy capture $E_{W,N}(\delta)$ with and without $2\times$ zeropadding for $N=128$ and window sizes $W=4$ and $W=5$. For almost all $\delta$, the non-zeropadded DFT (solid curves) achieves equal or higher cumulative energy capture than the $2\times$-zeropadded case (dashed) and when zeropadding does provide a slight advantage, the gain is marginal.

\noindent
{\bf Zeropadding in a noise-limited regime:} We must now account for noise coloring when quantifying signal energy capture.   Comparing the SNR of a whitened matched filter in the beamspace regime (ignoring interference) against the SNR in antenna space, we obtain
\begin{equation} \label{signal_energy_noiselimited_zp}
\eta(W) \;=\; 
\frac{\mathbf u_1^H \mathbf R_N^{-1}\mathbf u_1}
{\mathbf v_1^H \mathbf v_1/(2\sigma^2)}.
\end{equation}
Here, the linear transformation corresponding to a zeropadded DFT followed by beamspace windowing is  
$
\mathbf T = \mathbf W_{1}\mathbf{F}_{N_{\mathrm{FFT}}}\mathbf{Z},
$
where $\mathbf{Z}$ denotes the operator that pads its input with zeros.
The
corresponding colored noise covariance matrix is given by $ \mathbf R_N = 2 \sigma^2 \mathbf T \mathbf T^H$.
The noise variance per dimension, $\sigma^2$, cancels out in (\ref{signal_energy_noiselimited_zp}) and is included only for clarity of exposition.

\begin{figure}[h]
\centering
\subfloat[]{%
  \includegraphics[width=0.48\linewidth]{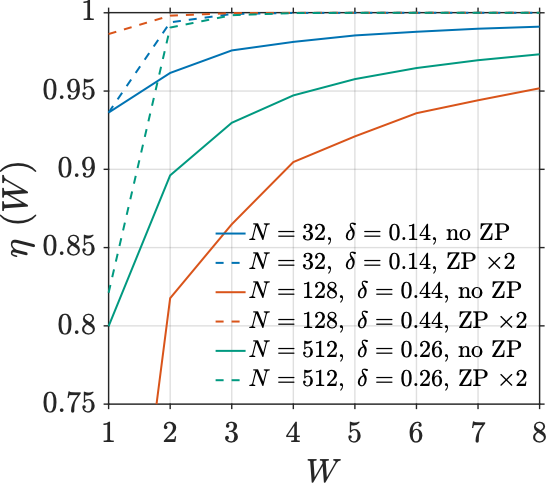}%
}\hspace{-0.2em}%
\subfloat[]{
  \includegraphics[width=0.48\linewidth]{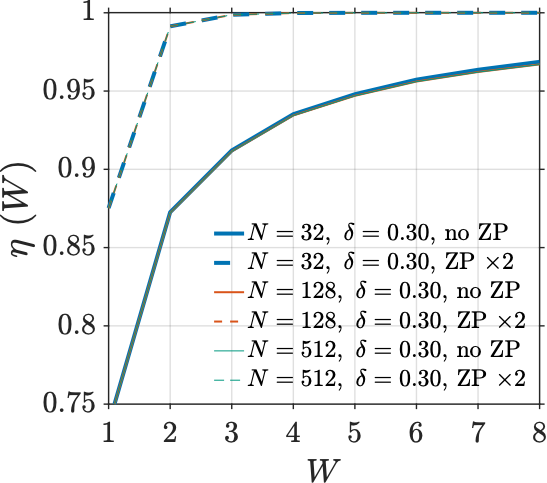}
}
\caption{Fractional energy capture $\eta(W)$ in a noise-limited regime with and without zeropadding. (a) $\theta^\circ=15$ for all array sizes considered.  (b) $\delta=0.3$ for all array sizes.}
\label{plt:desiredenergycapturewithnoise}
\end{figure}

As shown in Fig. \ref{plt:desiredenergycapturewithnoise}, the curves of fractional energy capture $\eta(W)$ grow rapidly with $W$, indicating efficient signal concentration in the noise-limited regime as well.  In this regime, zeropadding yields extremely efficient concentration for off-grid spatial frequencies, and is therefore expected to be beneficial in scenarios where interference is weaker and/or SNR is lower.  We defer quantitative comparisons to Section \ref{sec:numerical_examples}.

\subsection{Interference Concentration} \label{sec:interference concentration}

We now study linear detection with beamspace dimension reduction in an interference-limited regime using the simplified model (\ref{simplifiedmodel}), reduced to the $W$-dimensional beamspace model (\ref{simplifiedmodel_beamspace}) for a designated user, say user 1. 
The windowed beamspace signatures are given by $\mathbf u_k=\mathbf T\mathbf v_k$, where $\mathbf T = \mathbf W_{\mathbf{1}}\mathbf{F}_{N_{\mathrm{FFT}}}\mathbf{Z}$ and $\mathbf{Z}$ reduces to the identity matrix when no zeropadding is used.

\begin{figure}[h]
\centering
\subfloat[]{%
  \includegraphics[width=0.51\linewidth]{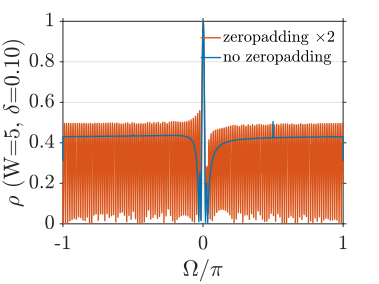}%
}\hspace{-1.6em}
\subfloat[]{
  \includegraphics[width=0.51\linewidth]{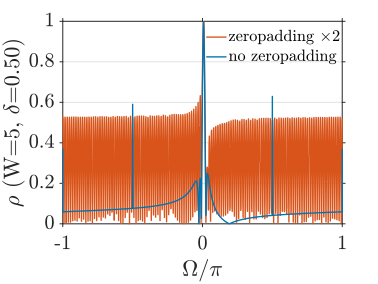}
}
\caption{Cosine similarities between the desired user's and the interferer's beamspace response, where the interferer's spatial frequency is swept in $[-\pi,\pi)$. For both figures, $N=128$, $W=5$. (a) The desired user's spatial frequency is at $\delta=0.1$ (b) at $\delta=0.5$, where $\delta$ is defined in the base $N$-point DFT grid.}
\label{plt:cosinesimilarity}
\end{figure}

The standard approach to developing geometric understanding of LMMSE detection is for systems in which the zero-forcing (ZF) solution exists. In such a setting, the LMMSE solution tends to the ZF correlator in an interference-limited regime \cite{madhow1994}. The ZF correlator is proportional to the projection of $\mathbf u_1$ orthogonal to the interference subspace spanned by the interference vectors,  $\mathcal S_I=\mathrm{span}\{\mathbf u_k:\,k\ge2\}$. It exists only if $\mathbf u_1\notin\mathcal S_I$. A necessary condition is that the dimension of $\mathcal S_I$ is smaller than the signal space dimension. 
The fraction of desired signal lost in nulling the $k$th interferer is $|\rho_k|^2$, where 
\begin{equation}
\rho_k \;=\; \frac{\big|\mathbf u_1^H \mathbf u_k\big|}{\|\mathbf u_1\|\,\|\mathbf u_k\|}.
\end{equation}
is the cosine similarity between the desired user and the $k$th interferer.  Under the worst-case assumption that the interferers are orthogonal to each other, the fraction of desired signal lost in nulling all interferers (numbered $2$ through $K$) is lower-bounded by $\sum_{k=2}^K |\rho_k|^2$.  

This worst-case approach to developing geometric understanding is not useful in reduced dimension beamspace: as shown in Fig. \ref{plt:cosinesimilarity}, the cosine similarities can be quite large, and the number of interferers is (much) larger than the signal space dimension. We must therefore look beyond geometric insights from zero-forcing in order to understand the efficacy of LMMSE reception in reduced-dimension beamspace.  Our approach is to demonstrate interference concentration by examining the eigenmodes of the {\it averaged} interference covariance matrix, and by deriving conservative design guidance from it.  We start with a general theorem stating that such averaged analysis is indeed conservative, and then specialize to the specific scenario of interest to us.

\begin{theorem}[Lower bound on expected SINR via mean interference covariance matrix]
\label{thm:meanSINRlower}
The average SINR (\ref{sinr}) for an LMMSE correlator, conditioned on the desired signal $\mathbf u_1$, and averaged across the interference and noise using a given probability law, can, in general, be lower bounded by replacing $\mathbf R_{I+N}$ by its expectation under that probability law:
\begin{equation*}
\mathbb E[\mathrm{SINR}] =P_1 \mathbb E\!\big[\mathbf u_1^{H}\mathbf R_{I+N}^{-1}\mathbf u_1\big] 
\;\ge\;
P_1 \mathbf u_1^{H}\,\mathbb E\!\big[\mathbf R_{I+N} \big]^{-1} \,\mathbf u_1.
\end{equation*}
\end{theorem} 
The proof, provided in Appendix \ref{app:proof_sinr_lower_bound}, draws on an operator Jensen's inequality. It is analogous to the observation that, for a random variable $X$ which is positive with probability one, $E[1/X] \geq 1/E[X]$ by the convexity of $1/x$ for $x > 0$.

We will apply this theorem to an interference-limited regime in which the number of users $K$ and the number of antennas $N$ grow, while keeping the size of the beamspace window used for per-user LMMSE fixed. The resource allocation strategy enforces a minimum spatial frequency separation between users scaling with the DFT bin size $2 \pi/N$, which allows $K$ to scale linearly with $N$.
We analyze the performance of the reduced-dimension LMMSE receiver from the perspective of the desired user, using window indices chosen as in Section \ref{sec:concentration}. 
The interferer spatial frequencies $\{\Omega_k\}_{k=2}^K$ are modeled as i.i.d. draws from a base distribution on $[-\pi,\pi)$, excluding a guard interval around the desired spatial frequency $\Omega_1$ imposed by the minimum spatial separation between users. 
For analytical simplicity, we do not enforce a minimum pairwise spacing among the interferers themselves. 
We further assume that the interferer powers $\{P_k\}_{k=2}^K$ are independent of their spatial frequencies.

The following proposition provides a simple expression for the expected interference correlation matrix $\overline{\mathbf R}_I$ in terms of the total interference power and an expectation over a single ``typical'' interferer.  We can then apply the lower bound in Theorem \ref{thm:meanSINRlower} to obtain geometric insights and design guidance.

\begin{proposition}[Mean interference + noise covariance]\label{prop:RbarI}
With the model above,
\begin{equation*}\label{R_bar}
\overline{\mathbf R}_{I + N}
:=\mathbb{E}\!\left[\mathbf R_{I + N} \right]
= P_{\mathrm{tot}}\;\mathbf M_{I} + 2\sigma^2\mathbf I,
\;
P_{\mathrm{tot}}=\sum_{k=2}^K P_k,
\end{equation*}
\begin{equation*} \label{M_I}
\mathbf M_I := \mathbb{E}\!\big[\mathbf u(\Omega)\,\mathbf u(\Omega)^{H}\big],
\end{equation*}
where the expectation defining $\mathbf M_I$ uses the interferer spatial frequency distribution restricted to the support outside the guard interval.
\end{proposition}

Combining the preceding proposition with Theorem \ref{thm:meanSINRlower}, we obtain the following lower bound on expected SINR:
\begin{equation}\label{sinr_lower_bound}
\begin{aligned}
\mathbb{E}[\mathrm{SINR}] \;\ge\; \underline{\mathrm{SINR}}
&= P_1\,\mathbf u_1^{H}\,\overline{\mathbf R}_{I+N}^{-1}\,\mathbf u_1 \\[2pt]
&= P_1\,\mathbf u_1^{H}\!\left(P_{\mathrm{tot}}\mathbf M_{I} + 2\sigma^2\mathbf I\right)\!\mathbf u_1 .
\end{aligned}
\end{equation}

\noindent{\bf SIR Margin:} We now introduce a simple metric summarizing the relative geometry of the signal and the interference. This is the average SIR for a hypothetical system with a single typical interferer whose power equals that of the desired user, which we term the ``SIR margin'':
\begin{equation} \label{sinr_margin}
\mathrm{SIR}_{\mathrm{margin}} := \mathbf u_1^{H}\,\mathbf M_I^{-1}\,\mathbf u_1.
\end{equation}
In an interference-limited regime, this quantity characterizes how large the total interference power can be relative to the desired user's power.  Using the lower bound 
$\underline{\mathrm{SINR}}$ as an approximation for the SINR, we obtain, ignoring noise, that 
\begin{equation} \label{sinr_approx}
\mathrm{SINR} \;\approx\; \underline{\mathrm{SIR}}
=  \frac{P_1}{P_{\mathrm{tot}}}\,\mathrm{SIR}_{\mathrm{margin}},
\end{equation}
Note that $P_1/P_{\mathrm{tot}}$ depends on both the number of interferers and their powers relative to that of the desired user. For equal-power users, (\ref{sinr_approx}) specializes to
\begin{equation} \label{SINR_K_dependence}
\mathrm{SINR} \approx  \frac{1}{K-1} \mathrm{SIR}_{\mathrm{margin}}.
\end{equation}
More generally, suppose that the distribution of relative powers between users is scaled such that $E[P_k] = 1$, with power control guaranteeing that $P_k \geq P_{min} > 0$ for all $k$.  Applying (\ref{sinr_approx}), we obtain the approximate worst-case prediction:
\begin{equation} \label{SINR_K_dependence2}
\mathrm{SINR}_{\rm min} \approx  \frac{P_{min}}{K-1} \mathrm{SIR}_{\mathrm{margin}}.
\end{equation}

Thus, $\mathrm{SIR}_{\mathrm{margin}}$ provides a practical tool for assessing how much interference power the system can tolerate, and thus for dimensioning the system (e.g., choosing user loading, guard intervals, and power control) in a high-SNR, interference-limited regime. After choosing these system parameters, the impact of noise can be accurately accounted for using the lower bound (\ref{sinr_lower_bound}), as illustrated by the numerical results in Section \ref{sec:numerical_examples} (see Table \ref{tab:sinr_pred_sim}).

\noindent
{\it Geometric insight:} Denoting by  $\{\lambda_i,\mathbf q_i\}$ the eigenvalue–eigenvector pairs for $\mathbf M_I$, we can rewrite the SIR margin as
\begin{equation} \label{sir_lower_bound2}
\mathrm{SIR}_{\mathrm{margin}} =  \sum_{i=1}^{W}\frac{\big|\mathbf q_i^{H}\mathbf u_1\big|^2}{\lambda_i}
\end{equation}
Large eigenvalues penalize alignment between the desired signal and the corresponding interference eigenvector, while small eigenvalues reward it. In the regime of interest, we find (see Fig. \ref{plt:eigpercentagewithdesireduserprojection} in Section \ref{sec:numerical_examples}) that $\mathbf M_I$ has only a few dominant eigenmodes {\it and} the desired signal retains energy along weaker interference modes. Thus, despite the high pairwise cosine similarities seen in Fig. \ref{plt:cosinesimilarity}, LMMSE reception is effective in attenuating interference without excessive loss in desired signal energy. 

\noindent
{\it Linear scaling of $K$ with $N$:} Our resource allocation strategy of maintaining a minimum spatial frequency separation scaling with the DFT bin size $\frac{2 \pi}{N}$ enables $K$ to scale linearly with $N$, and our numerical results show that these design guidelines yield excellent performance (see Section \ref{sec:numerical_examples}). From (\ref{SINR_K_dependence}), we see that linear scaling of $K$ with $N$ while maintaining high SINR requires that  $\mathrm{SIR}_{\mathrm{margin}}$ scale up at least as fast as $N$. 
This is indeed true, as stated in the following theorem proved in Appendix \ref{app:proof_sir_scaling}.
\begin{theorem}[Linear Scaling of $\mathrm{SIR}_{\mathrm{margin}}$] \label{thm:sinr_scaling}
Consider a beamspace window of size $W \geq 2$. 
Assuming interfering users are uniformly distributed in spatial frequency outside an guard interval proportional to $1/N$, $\mathrm{SIR}_{\mathrm{margin}}$ for the LMMSE receiver scales at least linearly with $N$.
\end{theorem}

Empirical evidence supporting this result is provided in Fig. \ref{plt:sinrmarginScaling}, which plots $\mathrm{SIR_\mathrm{margin}}$ versus $N$ for fixed $W=5$ on a log scale.

\section{Numerical Examples} \label{sec:numerical_examples}
In this section, we  illustrate the analytical framework developed in Section \ref{sec:concentration} and its design implications via numerical examples.
We consider LoS steering vectors on a ULA with half-wavelength spacing; array sizes $N \in \{32,64,128,256\}$; window sizes $W \in \{4,5\}$; and DFT sizes $N_{\mathrm{FFT}} \in \{N,2N\}$. Guard intervals are measured in bins on the base $N$-point DFT grid: an $x$-bin guard interval means that for any two users with spatial frequencies $\Omega_1$ and $\Omega_2$, we enforce
\begin{equation}
|\Omega_1 - \Omega_2| > x\,\frac{2\pi}{N}.
\end{equation}
The expectation (\ref{M_I}) defining the matrix $\mathbf M_I$ is computed by Monte Carlo simulations.

\begin{figure}[h]
\centering
\includegraphics[width=0.8\linewidth]{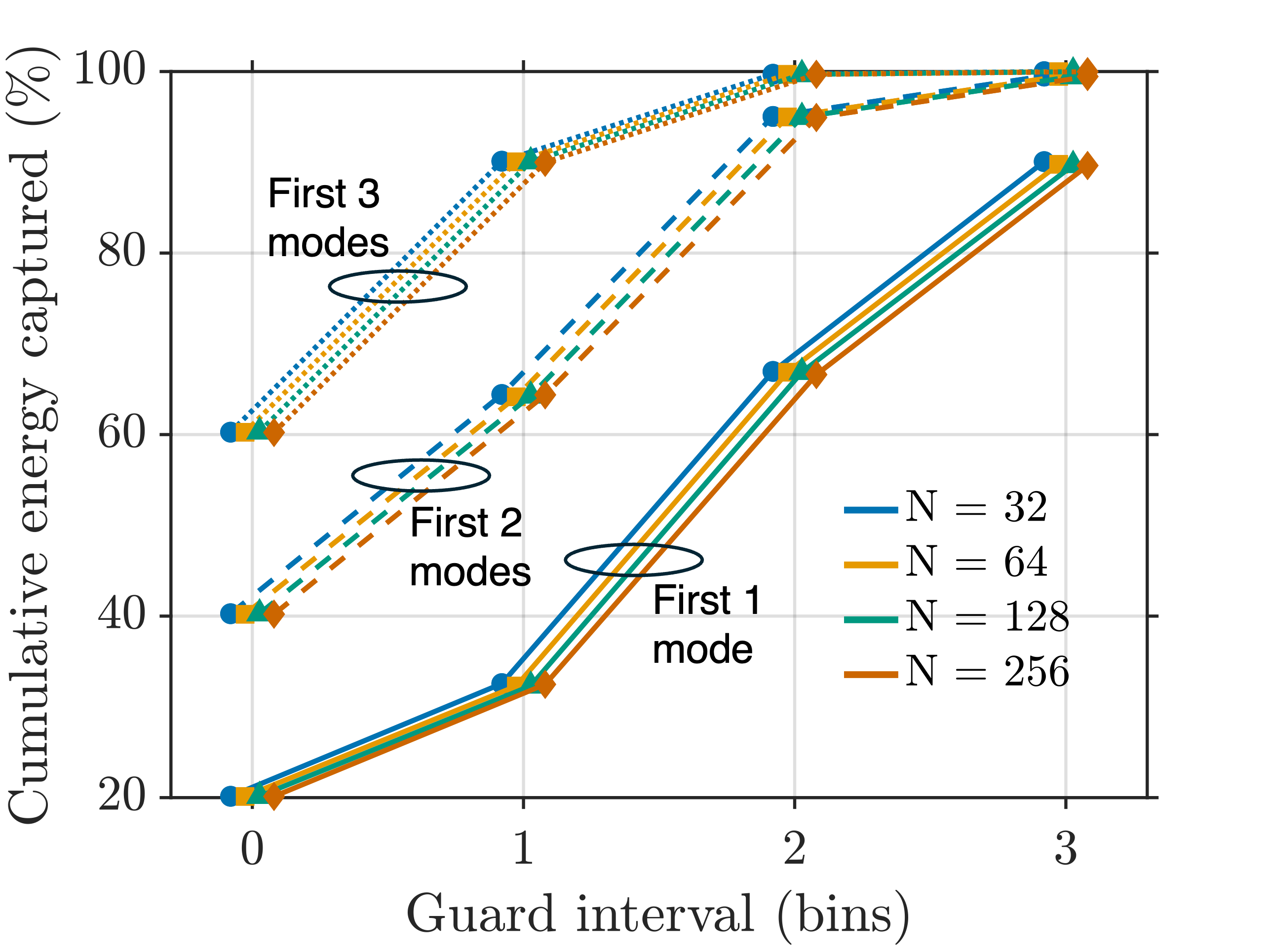}
\caption{Cumulative interference energy captured by the eigenvalues of 
$\mathbf M_I$ as a function of the guard interval for different array sizes, where we fix $W=5$ for each array size. We fix an off-grid offset $\delta=0.25$ for the desired user. Curves show the cumulative share of energy captured by the first $i$ modes ($i=1,\ldots,W-2$), where the eigenvalues are ordered from the largest to the smallest. A guard interval of 0 means that the only restriction on interferer placement is that their spatial frequencies are distinct from that of the desired user.}
\label{plt:eigpercentage}
\end{figure}

Interference concentration is characterized via a spectral analysis of $\mathbf M_I$. 
Fig. \ref{plt:eigpercentage} plots the cumulative fraction of interference energy captured by the leading eigenmodes, and shows that interference concentration is insensitive to array size $N$: the curves for different $N$ are nearly identical, with differences around $0.1\%$. 
As the guard interval increases, cumulative capture improves monotonically. With a 2-bin guard, about $67\%$ of the energy sits in the largest eigenvalue and about $95\%$ in the top two. A 3-bin guard interval (which corresponds to fewer simultaneous users) reaches around $90\%$ with one eigenmode alone. 

\begin{figure}[h]
\centering
\includegraphics[width=0.8\linewidth]{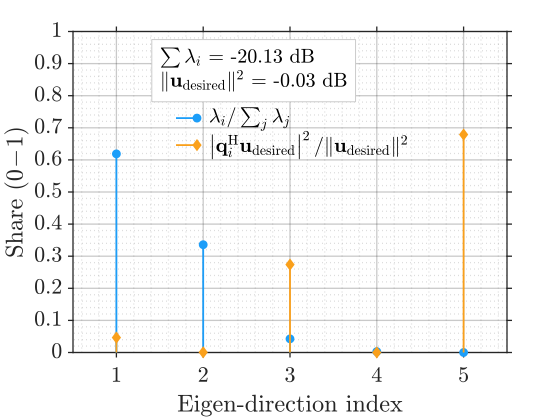}
\caption{Eigenvalue concentration and desired user projections where $N=128$, $W=5$  and $\delta=0.1$ and the guard interval around the desired user is 2 DFT bins. For this setting, the average interference power in the desired user's window is -20.13 dB, and the desired user loses a very small fraction of its energy due to windowing.}
\label{plt:eigpercentagewithdesireduserprojection}
\end{figure}

Fig.\ref{plt:eigpercentagewithdesireduserprojection}  illustrates the typical geometric relationship between the desired signal and the interference,  showing the relative strengths of the interference eigenvalues and the relative distribution of the desired user's energy among the interference eigenmodes.  We expect a high value of $\mathrm{SIR}_{\mathrm{margin}}$ in this setting because (1) the desired energy is concentrated on the weaker interference modes, hence strong modes can be suppressed with minimal desired signal loss; (2) the sum of the eigenvalues of $\mathbf M_I$ is about -20 dB, demonstrating the significant attenuation of the power of a typical interferer within the desired user's beamspace window.

\begin{figure}[h]
\setlength{\tabcolsep}{0pt} 
\centering
\includegraphics[width=0.7\linewidth]{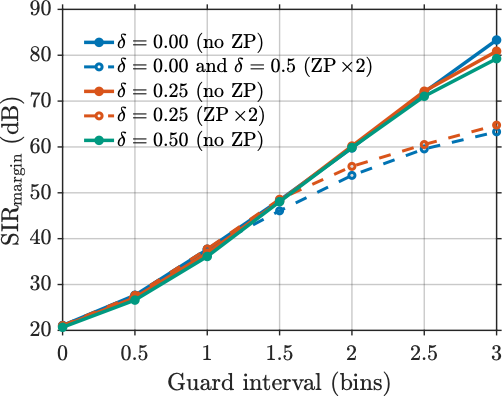} \caption{$\mathrm{SIR}_{\mathrm{margin}}$ versus guard interval for 
$N=128$ and $W=5$, comparing no-zeropadding ($N_{\mathrm{FFT}}=N$) and 
zeropadding $\times 2$ ($N_{\mathrm{FFT}}=2N$). 
Results are shown for three fractional offsets 
$\delta\in\{0,0.25,0.5\}$.}
\label{plt:sinrmarginDelta}
\end{figure}

Fig. \ref{plt:sinrmarginDelta} illustrates that $\mathrm{SIR}_{\mathrm{margin}}$ is only weakly sensitive to the desired user’s fractional DFT offset $\delta$, while improving with increase in guard interval (i.e., decrease in system loading). 
Comparing no-zeropadding with $2\times$ zeropadding, the curves coincide at $\delta=0$ (on-grid) and $\delta=0.5$ (half-bin), while for all other offsets zeropadding does not improve $\mathrm{SIR}_{\mathrm{margin}}$.

\begin{figure}[htb!]
\setlength{\tabcolsep}{0pt} 
\centering
\includegraphics[width=0.8\linewidth]{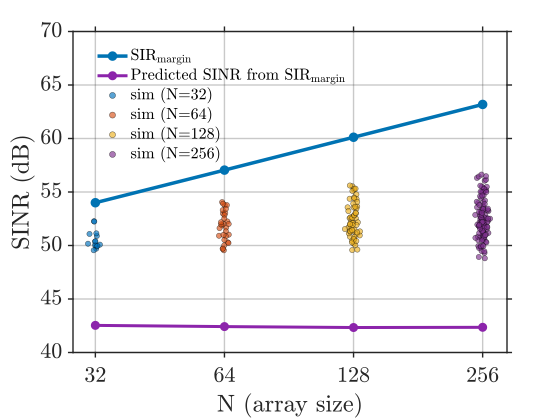} \caption{$\mathrm{SIR}_{\mathrm{margin}}$, predicted SINR from $\mathrm{SIR}_{\mathrm{margin}}$ and SINR values of simulated equal-power users plotted with respect to $10\log_{10}N$. The SNR per antenna is fixed to 40 dB for each simulation. For the simulated users, a guard interval of at least 2 DFT bins are used, therefore the number of users that can be fitted are different for each array size.}
\label{plt:sinrmarginScaling}
\end{figure}

We now investigate how $\mathrm{SIR}_{\mathrm{margin}}$ varies with array size $N$ for fixed $W=5$, with users spaced by at least 2 DFT bins, which corresponds to a system loading of $K \approx N/2$.  
Fig. \ref{plt:sinrmarginScaling} shows that $\mathrm{SIR}_{\mathrm{margin}}$ grows linearly with $N\in\{32,64,128,256\}$ (note that both axes are on a logarithmic scale). Combining with \eqref{SINR_K_dependence}, we can infer, for example, that we can indeed support linear growth of $K$ with $N$ while maintaining a target SINR.  Fig. \ref{plt:sinrmarginScaling} overlays simulated SINR values for different values of $N$.  While these exhibit some scatter across users, they consistently lie above the prediction $\mathrm{SINR}_{\mathrm{pred}} = \mathrm{SIR}_{\mathrm{margin}}(W)-10\log_{10}(K-1)$, indicating that \eqref{SINR_K_dependence} can provide a basis for conservative system sizing. By virtue of \eqref{SINR_K_dependence2}, similar design guidelines apply to systems with imperfect power control.

We now present simulation results showing that the preceding design guidelines indeed lead to predictable performance. 
In all examples we fix $N = 128$ antennas, use a beamspace window size $W = 5$, and apply a scheduling rule that enforces a two-DFT-bin separation between users in the $N_{\mathrm{FFT}} = N$ base frame, considering both the no-zeropadding case and the $2\times$ zeropadding case. 
The average interference covariance $\mathbf{M}_I$ is computed for a desired user with fractional offset $\delta = 0.25$ and a two-DFT-bin separation from its interferers; exploiting the observed insensitivity of $\mathrm{SIR}_{\text{margin}}$ to $\delta$, we use this same averaged $\mathbf{M}_I$ to approximate the margin for all users.

\begin{figure}[h]
\centering
\subfloat[60 dB beamformed receive SNR per user.]{
\includegraphics[width=0.8\linewidth]{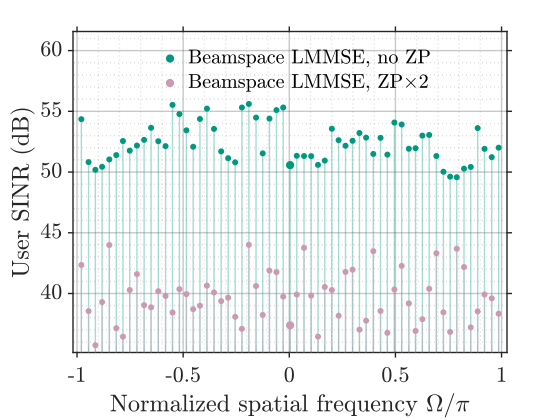}
}\par\vspace{-1.5em}  
\subfloat[10 dB beamformed receive SNR per user.]{
\includegraphics[width=0.8\linewidth]{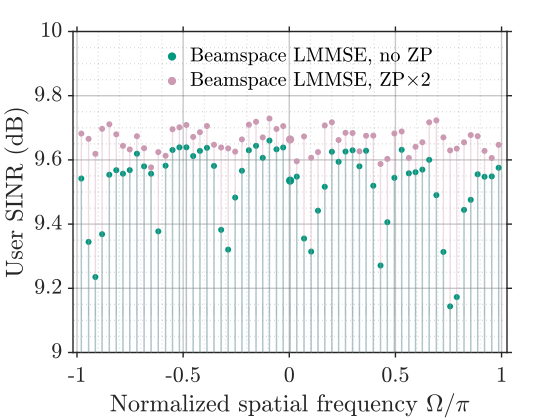}}
\caption{$\mathrm{SINR}$ after LMMSE detection for each user, with users separated by at least 2 DFT bins in the spatial frequency domain, and for the two plots, the user spatial frequencies are the same.}
\label{plt:SINRsims}
\end{figure}

For a fixed spatial-frequency distribution (the same for all experiments, with $K=61$ users), we consider both low- and high-noise regimes. We consider equal power settings, varying the beamformed SNR per user from 10 dB (noise-limited) to 60 dB (interference-limited). We also consider a scenario with power disparities, with half the users at beamformed SNR of 10 dB and half at 20 dB. We summarize the SINR predictions obtained from \eqref{sinr_lower_bound} together with the simulated minimum and mean SINR values in Table \ref{tab:sinr_pred_sim}. Full SINR simulation results for equal power users contrasting interference-limited (60 dB beamformed SNR) and noise-limited (10 dB beamformed SNR) regimes are plotted in Fig. \ref{plt:SINRsims}. 

For equal powers, we predict an interference-limited SINR of about $42$ dB:
\begin{equation}
\mathrm{SINR}_{\text{dB}} \;\approx\; \mathrm{SIR}_{\text{margin,dB}} - 10\log_{10}(K-1)
\;\approx\; 42~\text{dB}.
\end{equation}
When the beamformed SNR is substantially smaller than this quantity, we expect little degradation due to interference.  As the beamformed SNR increases, we expect the SINR to saturate. 
The specific predictions at each SNR are given by \eqref{sinr_approx}, the lower bound on expected SINR. Table \ref{tab:sinr_pred_sim} shows that this prediction is consistently smaller than the mean SINR, and is close to the minimum SINR, both without zeropadding and with $2\times$ zeropadding.  
We note that zeropadding provides little benefit, with marginal SINR improvement in noise-limited regimes, and slightly poorer performance in interference-limited regimes. 

\begin{table}[h]\label{SINRpredSINRsim}
    \centering
    \small 
    \setlength{\tabcolsep}{2pt} 
    \caption{Predicted and simulated SINR for different beamformed SNR and interferer-power configurations with $N=128$, $W=5$ and a guard interval of 2 DFT bins.}
    \label{tab:sinr_pred_sim}
    \begin{tabular}{@{}p{0.42\columnwidth}lcc@{}} 
        \toprule
        SNR configuration 
            & Statistic 
            & no ZP (dB) 
            & ZP $2\times$ (dB) \\
        \midrule
        \multirow{3}{=}{10 dB, equal power}
            & prediction & 9.43 & 9.58 \\
            & sim (min)  & 9.14 & 9.58 \\
            & sim (mean) & 9.52 & 9.66 \\
        \midrule
        \multirow{3}{=}{10 dB, half of the interferers have 10 dB more}
            & prediction & 8.90 & 9.07 \\
            & sim (min)  & 8.66 & 9.04 \\
            & sim (mean) & 9.03 & 9.21 \\
        \midrule
        \multirow{3}{=}{20 dB, half of the interferers have $10$ dB less}
            & prediction & 18.90 & 19.07\\
            & sim (min)  & 18.59 & 19.01 \\
            & sim (mean) & 19.06 & 19.21 \\
        \midrule
        \multirow{3}{=}{30 dB, equal power}
            & prediction & 28.19 & 27.99 \\
            & sim (min)  & 28.25 & 26.99 \\
            & sim (mean) & 28.33 & 27.59 \\
        \midrule
        \multirow{3}{=}{60 dB, equal power}
            & prediction & 42.23 & 37.91 \\
            & sim (min)  & 49.57 & 35.74 \\
            & sim (mean) & 52.71 & 40.19 \\
        \bottomrule
    \end{tabular}
\end{table}

\section{Wideband Regimes and Information-Theoretic Benchmarks}
\begin{figure}[htb!]
\setlength{\tabcolsep}{0pt} 
\centering
\includegraphics[width=0.8\linewidth]{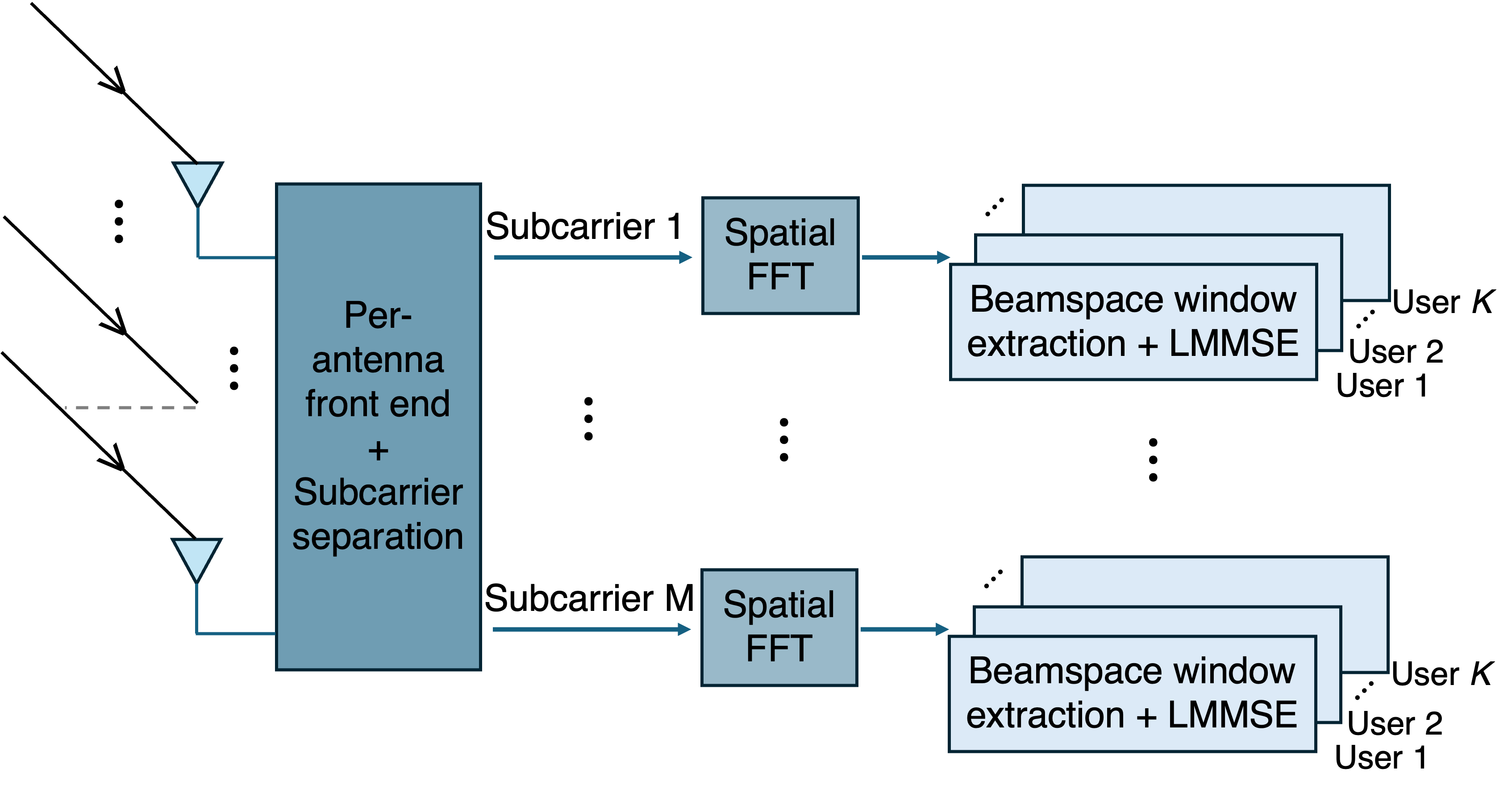} \caption{MIMO-OFDM with subcarrier beamspace processing system model.}
\label{plt:MIMOOFDMSystemModel}
\end{figure}

We now extend the geometric intuitions developed in the previous sections for simplified narrowband models to wideband multiuser channels. 
Fig. \ref{sparsityWRTfrequency} shows that for large fractional bandwidths, sparsity in beamspace is maintained over small subbands. 
This motivates investigation of fundamental limits for a MIMO-OFDM architecture with per-subcarrier beamspace LMMSE processing, illustrated in Fig. \ref{plt:MIMOOFDMSystemModel}. 
This architecture parallelizes low-dimensional LMMSE processing across users and subcarriers.

\noindent
{\it Guard interval scheduling in wideband beamspace:} To leverage the interference-concentration property in each subband, we employ a spatial-frequency scheduling rule that assigns a guard interval in spatial frequency around each user’s most significant (typically LoS) path. 
Because the spatial frequency depends on $f$, the spatial–frequency separation between two rays also changes across the band. 
For two rays with reference spatial frequencies $\Omega_1^{\mathrm{ref}}$ and $\Omega_2^{\mathrm{ref}}$, we have
\begin{equation}
\Delta\Omega(f) = |\Omega_1(f) - \Omega_2(f)| 
= \big|\Omega_1^{\mathrm{ref}} - \Omega_2^{\mathrm{ref}}\big|\left(1 + \frac{f}{f_c}\right),
\end{equation}
which is increasing in $f$ and does not vanish for distinct rays within the bandwidth. 
Therefore, to ensure that a minimum spatial–frequency separation holds over the entire band, we enforce the guard intervals at the lowest frequency.
In addition, to prevent two rays from becoming closer than the guard interval in any subband due to $2\pi$ wrapping, we require $|\Omega(B/2)| < \pi$, which yields
\begin{equation}
\left|\,1+\frac{B}{2f_c}\right|\sin(\theta) < 1
\quad \Rightarrow \quad
|\theta| < \sin^{-1}\!\left(\frac{1}{1+\frac{B}{2f_c}}\right).
\end{equation}
This condition limits the field of view. However, for a 20\% fractional bandwidth, the bound is $\sin^{-1}(1/1.1)\approx 65^\circ$, which is wider than the typical operational field of view $[-60^\circ,60^\circ]$, and thus does not introduce an additional limitation for practical systems. 

\noindent
{\it Wideband MIMO channel construction:} In order to benchmark the spectral-efficiency performance of the proposed system, we construct frequency-selective MIMO channel matrices from real-world measurements at 28.5 GHz in \cite{charbonnier2020calibration}. 
For each realization, we sample $K$ users from the dataset and form the channel matrix
\begin{equation}
\mathbf{H}(f) \in \mathbb{C}^{N \times K}, \quad
\mathbf{H}(f) = \big[\,\mathbf{h}_1(f)\ \cdots\ \mathbf{h}_K(f)\,\big],
\end{equation}
where $\mathbf{h}_k(f)$ is given by \eqref{eq:ChannelFreqSum}. For each user, we normalize the transmit power so that the beamformed SNR of the dominant path is the same across users. The powers of the remaining paths follow the relative path losses in the dataset.

\noindent
{\it Benchmarks:} Assuming Gaussian inputs, the spectral efficiency of the resulting frequency-selective MIMO channel, averaged over users and frequency, is
\begin{equation}
C_{\text{unconstrained}}
= \frac{1}{K B} \int_{B} \log_2 \det\!\left(
\mathbf{I} + \frac{\mathbf{H}(f)\mathbf{H}^H(f)}{2\sigma^2}
\right) df.
\label{spectralEfficiency}
\end{equation}
This benchmark places no constraints on implementation complexity, so we refer it as the unconstrained-complexity spectral efficiency.

Similar to (\ref{simplifiedmodel}), the MIMO-OFDM system in Fig. \ref{plt:MIMOOFDMSystemModel} is modeled, on each frequency $f$, as
\begin{equation}
\mathbf y(f) \;=\; \mathbf h_1(f) \;+\; \mathbf I_k(f) \;+\; \mathbf n,
\end{equation}
where $\mathbf I_k(f)=\sum_{k=2}^{K} \mathbf h_k(f)$, labeling the desired user as $1$ and the interferers starting from $2$.
The user symbols are unit-energy and uncorrelated across users, and the noise is white and Gaussian, $\mathbf n \sim \mathcal{CN}\big(\mathbf 0,\,2\sigma^2 \mathbf I\big)$.
The interference-plus-noise covariance matrix therefore is
\begin{equation}
\mathbf R_{I+N}(f) =
\mathbf R_I(f) + 2\sigma^2 \mathbf I,
\end{equation}
where $\mathbf R_I(f) = \mathbb{E}[\mathbf I_1(f)\mathbf I_1^H(f)]$.

For per-user, per-subcarrier LMMSE reception, we obtain SINR as a function of frequency as follows:
\begin{equation}
\mathrm{SINR}(f) \;=\; \mathbf h_1^H(f)\, \mathbf R_{I+N}^{-1}\, \mathbf h_1(f).
\label{capacity_SINR}
\end{equation}
Beamspace dimension reduction applies the linear transform $\mathbf T = \mathbf W_{\mathbf{1}}\mathbf{F}_{N_{\mathrm{FFT}}}$, yielding
\begin{equation}
\mathbf W_{\mathbf{1}}\mathbf{F}_{N_{\mathrm{FFT}}}{\mathbf{y}}(f)=\tilde{\mathbf{y}}(f) \;=\; \tilde{\mathbf h}_1(f) \;+\; \tilde{\mathbf I}_k(f) \;+\; \tilde{\mathbf n}.
\end{equation}
The corresponding SINR is
\begin{equation}
\mathrm{SINR}(f) \;=\; \tilde{\mathbf h}_1^H(f)\, \tilde{\mathbf R}_{I+N}^{-1}\, \tilde{\mathbf h}_1(f).
\label{capacity_SINR}
\end{equation}
The spectral efficiency averaged over bandwidth and users can now be bounded below as
\begin{equation}
C \;=\; \frac{1}{B}\int_{B} \frac{1}{K}\sum_{k=1}^{K} \log_2\!\big(1+\mathrm{SINR}(f)\big)\, df.
\label{capacity}
\end{equation}

\noindent
{\it Simulation model:} The channel for each simultaneous user in our simulations consists of $24$–$36$ paths drawn from the measured data in \cite{charbonnier2020calibration}, with fractional bandwidth set at 20\%. The number of simultaneous users is set to $K=16$: we impose a guard interval of at least $0.95$ of a DFT bin between the dominant paths of different users in the minimum frequency of the bandwidth, which limits the number of simultaneous users that we can generate from the dataset in \cite{charbonnier2020calibration}. We set $N=32$ to maintain a 50\% load factor as in the simplified model in Sections \ref{sec:concentration} and \ref{sec:numerical_examples}.  We consider per-subcarrier LMMSE reception after beamspace dimension reduction with window size $W=5$. In order to separate out the effects of multipath propagation from beam squint, we also consider a setting in which only the dominant path for each user is considered.

\begin{figure}[h]
\centering
\subfloat[All multipaths included in the channel.]{
\includegraphics[width=0.9\linewidth]{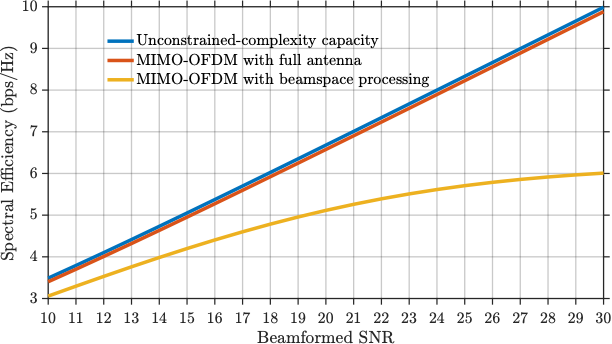}
}\par
\subfloat[Only the most dominant path of each user are considered.]{
\includegraphics[width=0.9\linewidth]{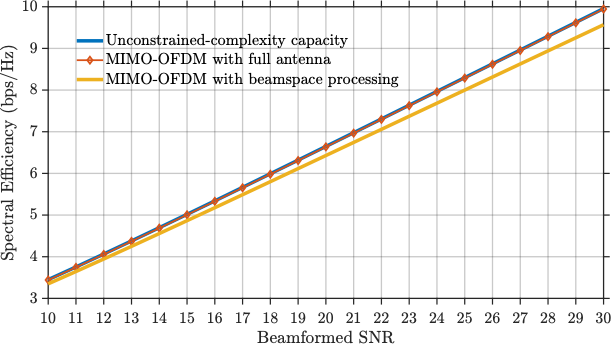}}
\caption{Spectral efficiency versus each user’s beamformed SNR.}
\label{plt:capacities}
\end{figure}

In Fig. \ref{plt:capacities}, we compare the spectral efficiency for our per-subcarrier dimension-reduced beamspace LMMSE scheme against two benchmarks: the unconstrained-complexity spectral efficiency in \eqref{spectralEfficiency} and per-subcarrier LMMSE reception with the full antenna array.  These two benchmarks closely track each other in both the multipath and dominant-path-only settings: in our 50\% loaded regime, linear interference suppression incurs little loss of optimality (less than $0.5$ dB over the range $10$–$30$ dB). On the other hand, our beamspace dimension-reduced scheme behaves very differently with and without multipath. It tracks the benchmarks closely when we only consider dominant paths, but its spectral efficiency saturates in a multipath setting, albeit at values which are high enough to support moderately large constellations.   

\begin{figure}[h]
\centering
\subfloat[Multipath setting.]{
  \includegraphics[width=0.7\linewidth]{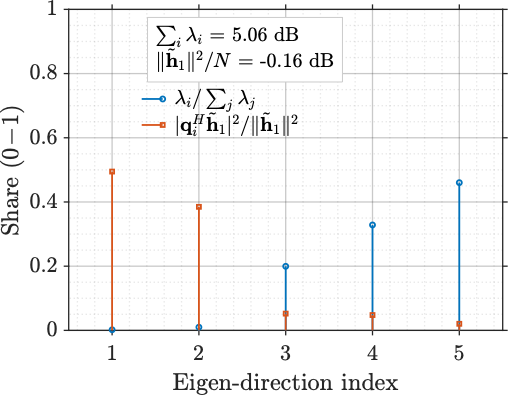}
}\par\vspace{-1em}
\subfloat[Dominant paths only.]{
\includegraphics[width=0.7\linewidth]{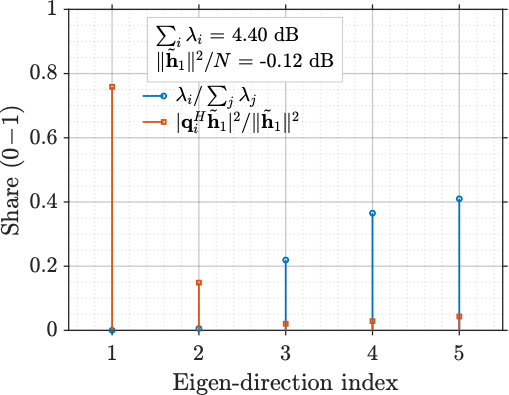}
}\par\vspace{-1em}
\subfloat[SIR along the subbands.]{
\includegraphics[width=0.7\linewidth]{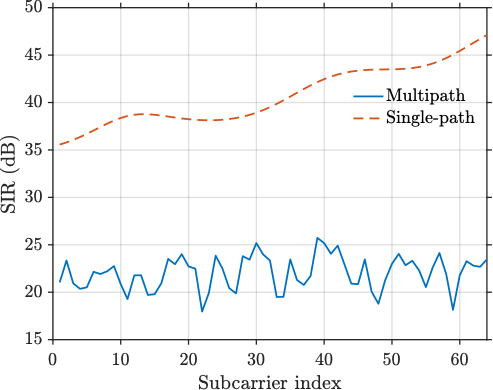}}
\caption{Eigenvalue concentration of the interference covariance matrix and projection of the desired-user channel onto the eigendirections at the carrier frequency (a-b). (c) SIR values with respect to subcarrier index for the desired user.}
\label{plt:eigenprojectionsRealData}
\end{figure}

In order to understand the difference in behavior of our scheme in the presence of multipath, recall that our scheduling rule enforces spatial separation for the dominant paths. In a multipath scenario, however, the secondary paths arrive at angles that are not subject to this constraint.
To highlight the resulting differences between single-path and multipath channels, Fig. \ref{plt:eigenprojectionsRealData}(a)–(b) plots, for a representative subcarrier and a representative user among the 16 simulated users, the eigenvalue shares of the interference covariance matrix along with the desired user’s energy on the same eigendirections.
For this user, the desired-channel norm in the window is slightly smaller and the interference energy slightly larger in the multipath setting. We find this trend common among users, but these differences alone are not large enough to explain the pronounced saturation in Fig. \ref{plt:capacities}(b). Instead, the key effect is how the desired user's energy is distributed relative to the dominant interference eigenmodes. 
With multipath, secondary interference paths can fall close to the spatial frequency of the desired user's dominant path, leading to greater alignment between the desired user and the dominant interference modes. Fig. \ref{plt:eigenprojectionsRealData}(c) quantifies this by tracking SIR across subcarriers. With $N=32$, $W=5$, and a one-bin guard interval, the SIR predicted using $\mathrm{SIR}_{\mathrm{margin}}$ is approximately $20$ dB. While the dominant-path-only model consistently exceeds this, the measured multipath channel collapses to an SIR floor of about $15-20$ dB. 

The performance saturation due to secondary paths not accounted for in resource allocation does not present a practical bottleneck for our mmWave dataset \cite{charbonnier2020calibration}, but these results highlight the importance of propagation modeling when evaluating our proposed architecture in other contexts (e.g., different frequency bands, different environments).

\section{Conclusion} \label{sec:conclusion}
We have provided analytical and geometric insights into why LMMSE with fixed beamspace window, and hence fixed per-user demodulation complexity, enables linear scaling of the number of users $K$ with number of antennas $N$.  Our lower bound on expected SINR provides explicit design guidance on resource allocation and power control. Such architectures extend naturally to wideband MU-MIMO OFDM via per-subcarrier, per-user LMMSE over small beamspace windows, but performance may saturate due to secondary paths not accounted for in resource allocation, with the extent depending on the propagation environment.
Our analysis also shows that the standard practice of zeropadding FFTs can lead to a performance drop in interference-limited regimes. 

We hope that the analytical insights presented here will motivate and guide practical DSP architectures for scaling massive multiuser MIMO.  While this work was initially motivated by sparse propagation in mmWave channels, it may be possible to exploit sparsity at lower carrier frequencies as well, especially in outdoor settings.  Of particular interest are the upper mid-band frequencies \cite{Upper_mid_band_NYU_journal}, the so-called FR3 band, being explored for NextG cellular systems, where the use of large digital arrays is anticipated.

\appendices
\section{Theorem on Energy Capture}
\label{app:proof_energy_capture}
\begin{proof}
We saw that the normalized DFT coefficients $y_n$ can be represented in terms of the Dirichlet kernel as in (\ref{normalized_DFT}).
The key idea is to lower bound the magnitude of the Dirichlet kernel by a sinc function using the inequality
\begin{equation}
|\sin(x)| \leq |x|.    
\end{equation}
We therefore obtain
\begin{equation} \label{dirichlet_sinc}
|D_N ( \omega )| = 
\big| \frac{\sin (N\omega/2)}{N \sin(\omega/2)} \big| \geq \big| \frac{\sin (N\omega/2)}{N \omega/2} \big|.
\end{equation}

For spatial frequency $\Omega \in [- \pi, \pi ]$, let us define a continuous DFT index
$\nu = \Omega/ (2 \pi/N) \in [-N/2,N/2]$, so that $\Omega = 2 \pi \nu/N$. The nearest DFT grid point can now be defined as
$n_0:=\mathrm{round}(\nu)\in\mathbb Z$, applying a modulo $N$ operation to bring it to the set
of indices $\{ -N/2,...,N/2-1 \}$ (for example, $n_0 = N/2$ is mapped to $n_0 = - N/2$).  
We have $\delta:=|\nu-n_0|\in[0,\tfrac12]$ as before, with $\nu=n_0 + \delta$ or $\nu=n_0 - \delta$.

The normalized DFT coefficients in (\ref{normalized_DFT}) can be rewritten as 
\begin{equation}
y_n = D_N (2 \pi (\nu - n)/N).    
\end{equation}

We can now apply (\ref{dirichlet_sinc}) with $\omega = 2 \pi (\nu - n)/N$ to obtain a lower bound on the fractional energy captured in the $n$th coefficient:
\begin{equation}
|y_n|^2\ \ge\ \sinc^2(n-\nu),
\end{equation}
where $\sinc(x) = \frac{\sin( \pi x)}{\pi x}$.
Let $\mathcal W^\star$ be the set of $W$ DFT indices for a user defined in Section \ref{sec:concentration}. Then we have
\begin{equation}
E_{W,N} (\delta) =\sum_{n\in\mathcal W^\star}\!|y_n|^2
\ \ge\
\sum_{n\in\mathcal W^\star} \sinc^2(n-\nu).
\end{equation}

Under $2\times$ zeropadding, the normalized $2N$-point DFT coefficients of the length $N$ array response can be written in terms of the length-$N$ Dirichlet kernel as
\begin{equation}
y_n^{(\mathrm{zp2})}
=
\frac{1}{\sqrt{2}}\,
D_N\!\left(
\frac{2\pi\nu}{N} - \frac{2\pi n}{2N}
\right).
\end{equation}
Taking magnitudes and using $|\sin(x)| \leq |x|$ gives
\begin{equation}
|y_n^{(\mathrm{zp2})}|
=
\frac{1}{\sqrt{2}}
\frac{|\sin(N\omega/2)|}{N\,|\sin(\omega/2)|}
\;\ge\;
\frac{1}{\sqrt{2}}
\frac{|\sin(N\omega/2)|}{N\,|\omega/2|},
\end{equation}
with
\begin{equation}
\omega
=
\frac{2\pi\nu}{N} - \frac{2\pi n}{2N}
=
\frac{2\pi}{N}\left(\nu - \frac{n}{2}\right).
\end{equation}
Thus
\begin{equation}
\frac{N\omega}{2}
=
\pi\!\left(\nu - \frac{n}{2}\right),
\end{equation}
and therefore
\begin{equation}
\frac{|\sin(N\omega/2)|}{N\,|\omega/2|}
=
\left|
\frac{\sin\!\left(\pi(\nu - n/2)\right)}{\pi(\nu - n/2)}
\right|
=
\sinc\!\left(\nu - \frac{n}{2}\right).
\end{equation}
Combining the inequalities yields
\begin{equation}
|y_n^{(\mathrm{zp2})}|^2
\;\ge\;
\frac{1}{2}\,
\sinc^2\!\left(\nu - \frac{n}{2}\right).
\end{equation}

Finally, summing over the optimally placed $W$-bin window $\mathcal W^\star$ in the $2N$-point DFT grid gives
\begin{equation}
E^{(\mathrm{zp2})}_{W,N}(\delta)
=
\sum_{n\in\mathcal W^\star} |y_n^{(\mathrm{zp2})}|^2
\;\ge\;
\frac{1}{2}
\sum_{n\in\mathcal W^\star}
\sinc^2\!\left(\nu - \frac{n}{2}\right).
\end{equation}
\end{proof}

\section{Theorem on Lower bound on Expected SINR}
\begin{proof}
For $\mathbf u \in \mathbb C^{W \times 1}$, and a random Hermitian positive definite matrix $\mathbf R_I$, we want to show that
\begin{equation}
\mathbb E\!\big[\mathbf u^{H}\mathbf R_I^{-1}\mathbf u\big]
\;\ge\;
\mathbf u^{H}\,\mathbb E\!\big[\mathbf R_I\big]^{-1}\,\mathbf u.
\end{equation}
We define $f(X) := X^{-1}$ on the set of $W\times W$ Hermitian positive definite matrices. 
This function is operator convex; see the block matrix argument in Bhatia \cite[§1.3]{Bhatia2007}.
By the Choi--Davis--Jensen (operator Jensen) inequality for unital positive linear maps (\!\!\! \cite{hansen1982jensen}) applied to the expectation, the matrix 
$\mathbb E[\mathbf R_I^{-1}] - (\mathbb E[\mathbf R_I])^{-1}$ is positive semidefinite.
Hence, for every $\mathbf u\in\mathbb C^W$, 
\begin{equation}
\mathbb E[\mathbf u^{H}\mathbf R_I^{-1}\mathbf u] - \mathbf u^{H}(\mathbb E[\mathbf R_I])^{-1}\mathbf u \ge 0,
\end{equation}
equivalently, 
\begin{equation}
\mathbb E[\mathbf u^{H}\mathbf R_I^{-1}\mathbf u] \ge \mathbf u^{H}(\mathbb E[\mathbf R_I])^{-1}\mathbf u.   
\end{equation}
\end{proof}

\label{app:proof_sinr_lower_bound}

\section{Theorem on Linear Scaling of SIR Margin with Number of Antennas}
\begin{proof}
We know that the SIR for the LMMSE receiver is lower bounded by that for any other linear receiver operating on the same signal space. Thus, while it is difficult to directly characterize the growth of $\mathrm{SIR}_{\mathrm{margin}}$ with $N$ for the LMMSE receiver, we are able to prove the result for a suboptimal ``matched filter'' style receiver. For the interference-limited scenario of interest, for any linear receiver $\mathbf c$, the SIR for the LMMSE receiver is lower bounded by
\begin{equation}
\mathrm{SINR} \geq \frac{P_1 |\mathbf c^H \mathbf u_1|^2}{\mathbf c^H \mathbf R_I \mathbf c}.    
\end{equation}

Fixing $\mathbf u_1$ and averaging across the interference locations, we obtain using Jensen's inequality as before that
\begin{equation} \label{sinr_lb_suboptimal}
\mathbb E[ \mathrm{SINR} ]\geq P_1 \frac{|\mathbf c^H \mathbf u_1|^2}{E[ \mathbf c^H \mathbf R_I \mathbf c ]} = \frac{P_1}{P_{tot}} \frac{|\mathbf c^H \mathbf u_1|^2}{E[|\mathbf c^H \mathbf u_2|^2]}.
\end{equation}
Without loss of generality, suppose that the desired user's spatial frequency $\Omega_1 \in [0, \frac{2 \pi}{N}]$, so that the FFT bins for $\Omega = 0$ and
$\Omega = \frac{2 \pi}{N}$ are included in the selected beamspace window (which is assumed to be of size $W \geq 2$).  We focus only on this 2 bin window for our lower bound.  For any $\Omega \in [- \pi, \pi]$, the response in the two selected FFT bins is given by
\begin{equation} \label{u_omega}
{\mathbf u} ( \Omega ) = [ D_N ( \Omega ) , D_N (\Omega - \frac{2 \pi}{N})]^T
\end{equation}
For $\Omega_1 \in [0 , 2 \pi/N]$, we obtain upon simplification that
\begin{equation} \label{u1_w2}
{\mathbf u}_1 = e^{j (N-1) \Omega/2} [ a(\Omega_1), b(\Omega_1) e^{j \pi/N}]^T
\end{equation}
where 
$a (\Omega_1) = \frac{\sin (N \Omega_1/2)}{N \sin (\Omega_1/2)} \geq 0$ and $b (\Omega_1) = \frac{\sin (N \Omega_1/2)}{N \sin ((2 \pi/N - \Omega_1)/2)} \geq 0$.
This motivates consideration of the ``matched filter'' style receiver 
\begin{equation} \label{mf}
\mathbf c = [1, e^{j \pi/N}]^T
\end{equation}
to gather the energy of the desired user in a manner that is independent of the specific realization of $\Omega_1 \in [0, 2 \pi /N]$. The key idea is that, while the desired signal contributes two terms of the same sign to the correlator output, by the properties of the Dirichlet kernel, interfering signals contribute terms of opposite sign, which yields non-trivial upper bounds on the interference contributions to the output.

Arguing as in the signal concentration discussion in Section \ref{sec:signal_concentration}, it is easy to obtain an $\Theta(1)$ lower bound on the desired signal's energy at the output, $|\mathbf c^H \mathbf u_1|^2$.  In order to show that the expected SINR scales as $N$, therefore, we see from (\ref{sinr_lb_suboptimal}) that it suffices to prove that $\mathbb E[|\mathbf c^H \mathbf u_2|^2]$ scales as $1/N$.  We assume that $\mathbf u_2 = \mathbf u (\Omega )$, where $\Omega$ is drawn uniformly from $[- \pi, \pi ]$ except for a guard interval $[- 0.5 \pi/N, 2.5 \pi/N ]$ around the desired signal's spatial frequency. This guarantees that $|\Omega - \Omega_1| \geq \frac{0.5 \pi}{N}$ regardless of the location of $\Omega_1$ in $[0, 2 \pi/N]$. From (\ref{u_omega}) and (\ref{mf}), we obtain upon simplification that
\begin{equation} \label{interference_term}
\begin{array}{l}
Z (\Omega ) = |\mathbf c^H \mathbf u_2| 
= \big\vert \frac{\sin (N \Omega/2 )}{N \sin (\Omega/2)} - \frac{\sin (N \Omega/2 )}{N \sin ((\Omega - 2 \pi/N)/2)} \big\vert \\
\\
~~~~~~~= \frac{|\sin (N \Omega/2 )|}{N} \big\vert \frac{\sin( \Omega/2 - \pi/N) - \sin (\Omega/2)}{\sin(\Omega/2) \sin (\Omega/2- \pi/N)} \big\vert.
\end{array}
\end{equation}
We wish to find an upper bound on $\mathbb E[Z^2]$ as we average over $\Omega$.   
We will show that $Z^2$ scales as $1/N^4$ for $\Omega$ ``far enough'' from the chosen 2-bin window.  For $\Omega$ ``close'' to the chosen window, $Z$ can be $\Theta(1)$, but despite such terms, the average value of $\mathbb E[Z^2]$ scales as $1/N$.  

For simplicity, we focus on obtaining upper bounds for $Z$ (and hence $Z^2$) for $\Omega \geq 2.5 \pi/N$.  The approach for analyzing $\Omega \leq - 0.5 \pi/N$ is entirely analogous, by symmetry relative to the chosen 2-bin window. Without loss of generality, assume that $\Omega$ is off-grid (i.e., it is not an integer multiple of $2 \pi/N$). This holds with probability one, and in any case, $Z=0$ for on-grid $\Omega$ not falling in the chosen beamspace window.  Let us first bound a term in the numerator in (\ref{interference_term}):
\begin{equation} \label{mvt}
|\sin( \Omega/2 - \pi/N) - \sin (\Omega/2)| = | \cos (\psi ) | \pi/N \leq \pi/N
\end{equation}
where the first equality, based on the mean value theorem, holds for some $\psi \in [\Omega/2 - \pi/N , \Omega/2 ]$. Furthermore, $|\sin ( N \Omega/2 )| \leq 1$. Plugging these into (\ref{interference_term}), we obtain
\begin{equation} \label{interference_term2}
Z (\Omega ) \leq \frac{\pi}{N^2} ~ \frac{1}{|\sin(\Omega/2) \sin (\Omega/2- \pi/N)|}.
\end{equation}
Now, consider ``far-away'' $\Omega$ satisfying $\Omega/2 - \pi/N \geq \alpha$, where $0< \alpha + \pi/N \leq \pi/2$.
Since $|\sin (x)|$ is an increasing function of $|x| \in [0, \pi/2]$, we have for such $\Omega$ that 
\begin{equation}
|\sin(\Omega/2) \sin (\Omega/2- \pi/N)| \geq \sin^2 \alpha.    
\end{equation}
Plugging into (\ref{interference_term2}), we obtain that
\begin{equation} \label{interference_term3}
Z (\Omega) \leq \frac{\pi}{N^2 \sin^2 \alpha} ~,~~ 2 \alpha + 2 \pi/N \leq \Omega \leq \pi
\end{equation}
Since $\alpha > 0$ is chosen independently of $N$, we see that the contribution to $\mathbb E[Z^2]$ from such $\Omega$ scales as $1/N^4$.  The same scaling also holds for $\Omega \in [- \pi, - 2 \alpha]$.

It remains to understand the contribution of $\Omega \in [2.5 \pi/N, 2 \alpha + 2 \pi/N)$ (and analogously, for $\Omega \in (- 2 \alpha , -0.5 \pi/N]$) to $\mathbb E[Z^2]$.
Setting $\Omega' = \Omega - 2 \pi/N \in [0.5 \pi/N , 2 \alpha )$, we have 
\begin{equation}
\begin{array}{l}
 |\sin(\Omega/2) \sin (\Omega/2- \pi/N)| \geq \sin^2 (\Omega'/2) \\
 ~~~~~\geq \frac{\Omega'^2}{4} \left( 1 - \Omega'^2/24 \right)^2 \geq \frac{\Omega'^2}{4} (1 - \alpha^2/6)^2
 \end{array}    
\end{equation}
using the fact that $\sin (x)$ is increasing for $x \in [0, \pi/2 )$, and that $ \sin (x) \geq x - x^3/3! = x(1 - x^2/6)$
 in this range.  Plugging into (\ref{interference_term2}), we obtain
 that
 \begin{equation} \label{interference_term4}
Z (\Omega ) \leq \frac{4\pi}{N^2 (1 - \alpha^2/6)^2} \frac{1}{\Omega'^2} ~,~~ 0.5 \pi/N \leq \Omega' \leq 2 \alpha
\end{equation}
The contribution of such $\Omega$ to $\mathbb E[Z^2]$ is therefore upper bounded by the following term:
 \begin{equation}  \label{interference_term5}
 \begin{array}{l}
 T \leq  \left( \frac{4\pi}{N^2 (1 - \alpha^2/6)^2} \right)^2 \frac{1}{2 \pi - 3 \pi/N} \int_{0.5 \pi/N}^{2 \alpha} \frac{1}{\Omega'^4} d\Omega' \\
~~  = \frac{48 \pi^2}{N^4 (1 - \alpha^2/6)^4 (2 \pi - 3 \pi/N)} \left( 1/(0.5 \pi/N)^3 - 1/(2 \alpha)^3 \right) \\
= \Theta (1/N).
\end{array}
 \end{equation}
 An analogous upper bound holds for $\Omega \in [- 2 \alpha , - 0.5 \pi/N]$.
 Putting these together, we find that the contribution of ``nearby'' interferers to $\mathbb E[Z^2]$ scales as $O(1/N)$, while that of ``far'' interferers scales as $O(1/N^4)$, for the suboptimal matched filter style receiver with a beamspace window of size 2. Since the desired user contribution scales as $\Theta (1)$, we conclude from (\ref{sinr_lb_suboptimal}) that, as $N$ gets large, the expected SIR for the receiver (\ref{mf}) satisfies
 \begin{equation}
 \mathbb E[ \mathrm{SIR} ]\geq \frac{P_1}{P_{tot}} c N
 \end{equation}
 for some constant $c$.  This implies that $\mathrm{SIR}_{\mathrm{margin}}$ for the LMMSE receiver scales at least as well as $\Theta (N)$.

\end{proof}
\label{app:proof_sir_scaling}

\section*{Acknowledgments }
This work was supported in part by the Center for Ubiquitous Connectivity (CUbiC), sponsored by Semiconductor Research Corporation (SRC) and Defense Advanced Research Projects Agency (DARPA) under the JUMP 2.0 program, and in part by the National Science Foundation under grant CNS-21633.

\bibliographystyle{IEEEtran}
\bibliography{ref}

@article{Upper_mid_band_NYU_journal,
  author={Kang, Seongjoon and Mezzavilla, Marco and Rangan, Sundeep and Madanayake, Arjuna and Venkatakrishnan, Satheesh Bojja and Hellbourg, Grégory and Ghosh, Monisha and Rahmani, Hamed and Dhananjay, Aditya},
  journal={IEEE Open Journal of the Communications Society}, 
  title={Cellular Wireless Networks in the Upper Mid-Band}, 
  year={2024},
  volume={5},
  number={},
  pages={2058-2075},
  doi={10.1109/OJCOMS.2024.3373368}}

@INPROCEEDINGS{Abdelghany20_scalable_nonlinear,
  author={Abdelghany, Mohammed and Rasekh, Maryam Eslami and Madhow, Upamanyu},
  booktitle={2020 IEEE 21st International Workshop on Signal Processing Advances in Wireless Communications (SPAWC)}, 
  title={Scalable Nonlinear Multiuser Detection for mmWave Massive {MIMO}}, 
  year={2020},
  volume={},
  number={},
  pages={1-5},
  doi={10.1109/SPAWC48557.2020.9154238}}

@INPROCEEDINGS{Abdelghany19_beamspace_precoding,
  author={Abdelghany, Mohammed and Madhow, Upamanyu and Tölli, Antti},
  booktitle={2019 53rd Asilomar Conference on Signals, Systems, and Computers}, 
  title={Efficient Beamspace Downlink Precoding for mmWave Massive {MIMO}}, 
  year={2019},
  volume={},
  number={},
  pages={1459-1464},
  doi={10.1109/IEEECONF44664.2019.9048656}}

@INPROCEEDINGS{castaneda2021,
  author={Castañeda, Oscar and Boynton, Zachariah and Mirfarshbafan, Seyed Hadi and Huang, Shimin and Ye, Jamie C. and Molnar, Alyosha and Studer, Christoph},
  booktitle={ESSCIRC 2021 - IEEE 47th European Solid State Circuits Conference (ESSCIRC)}, 
  title={A Resolution-Adaptive 8 mm2 9.98 Gb/s 39.7 pJ/b 32-Antenna All-Digital Spatial Equalizer for mmWave Massive {MU-MIMO} in 65nm {CMOS}}, 
  year={2021},
  volume={},
  number={},
  pages={247-250},
  doi={10.1109/ESSCIRC53450.2021.9567843}}

@ARTICLE{mirfarshbafan2020,
  author={Mirfarshbafan, Seyed Hadi and Gallyas-Sanhueza, Alexandra and Ghods, Ramina and Studer, Christoph},
  journal={IEEE Transactions on Circuits and Systems I: Regular Papers}, 
  title={Beamspace Channel Estimation for Massive {MIMO} mmWave Systems: Algorithm and {VLSI} Design}, 
  year={2020},
  volume={67},
  number={12},
  pages={5482-5495},
  doi={10.1109/TCSI.2020.3023023}}

@INPROCEEDINGS{delafrooz2025,
  author={Oveys Delafrooz and Jiyoon Han and Wei Tang and Zhengya Zhang and Upamanyu Madhow},
  booktitle={2025 61st Annual Allerton Conference on Communication, Control, and Computing}, 
  title={Scaling Wideband Massive {MIMO} Radar via Beamspace Dimension Reduction}, 
  year={2025},
  volume={},
  number={},
  pages={},
  }

@INPROCEEDINGS{cebeci2024,
  author={Cebeci, Canan and Noroozi, Oveys Delafrooz and Madhow, Upamanyu},
  booktitle={2024 58th Asilomar Conference on Signals, Systems, and Computers}, 
  title={Scaling mmWave {MU-MIMO}: Information-Theoretic Guidance Using Real-World Data}, 
  year={2024},
  volume={},
  number={},
  pages={1620-1624},
  doi={10.1109/IEEECONF60004.2024.10942628}}

@article{chapman1976partial,
  title={Partial adaptivity for the large array},
  author={Chapman, DEANJ},
  journal={IEEE Transactions on Antennas and Propagation},
  volume={24},
  number={5},
  pages={685--696},
  year={1976},
  publisher={IEEE}
}

@ARTICLE{madhow1994,
  author={Madhow, U. and Honig, M.L.},
  journal={IEEE Transactions on Communications}, 
  title={{MMSE} interference suppression for direct-sequence spread-spectrum {CDMA}}, 
  year={1994},
  volume={42},
  number={12},
  pages={3178-3188},
  doi={10.1109/26.339839}}

@INPROCEEDINGS{tiledbeamspace,
  author={Han, Jiyoon and Cebeci, Canan and Tang, Wei and Zhang, Zhengya and Madhow, Upamanyu},
  booktitle={2024 IEEE 100th Vehicular Technology Conference (VTC2024-Fall)}, 
  title={Tiled Beamspace Processing for Scaling mmWave Massive {MU-MIMO}}, 
  year={2024},
  volume={},
  number={},
  pages={1-6},
  doi={10.1109/VTC2024-Fall63153.2024.10757813}}

@ARTICLE{reed1974adaptivearrays,
  author={Reed, I.S. and Mallett, J.D. and Brennan, L.E.},
  journal={IEEE Transactions on Aerospace and Electronic Systems}, 
  title={Rapid Convergence Rate in Adaptive Arrays}, 
  year={1974},
  volume={AES-10},
  number={6},
  pages={853-863},
  doi={10.1109/TAES.1974.307893}}

@INPROCEEDINGS{MUSICwithbeamspace,
  author={Zoltowski, M.D. and Kautz, G.M. and Silverstein, S.D.},
  booktitle={Fifth ASSP Workshop on Spectrum Estimation and Modeling}, 
  title={Simultaneous sector processing via {ROOT-MUSIC} for large sensor arrays}, 
  year={1990},
  volume={},
  number={},
  pages={372-376},
  doi={10.1109/SPECT.1990.205522}}

@ARTICLE{ESPIRITtwithbeamspace,
  author={Guanghan Xu and Silverstein, S.D. and Roy, R.H. and Kailath, T.},
  journal={IEEE Transactions on Signal Processing}, 
  title={Beamspace {ESPRIT}}, 
  year={1994},
  volume={42},
  number={2},
  pages={349-356},
  doi={10.1109/78.275607}}

@techreport{Ward1994STAP,
  author       = {James Ward},
  title        = {Space-Time Adaptive Processing for Airborne Radar},
  institution  = {MIT Lincoln Laboratory},
  type         = {Technical Report},
  number       = {TR-1015},
  address      = {Lexington, MA},
  year         = {1994},
  month        = dec,
  note         = {Also available via DTIC as ADA293032},
  url          = {https://apps.dtic.mil/sti/tr/pdf/ADA293032.pdf}
}

@INPROCEEDINGS{CapMIMO,
  author={Sayeed, Akbar and Behdad, Nader},
  booktitle={2010 48th Annual Allerton Conference on Communication, Control, and Computing (Allerton)}, 
  title={Continuous aperture phased {MIMO}: Basic theory and applications}, 
  year={2010},
  volume={},
  number={},
  pages={1196-1203},
  doi={10.1109/ALLERTON.2010.5707050}}

@INPROCEEDINGS{BradyAnalog,
  author={Brady, John H. and Sayeed, Akbar M.},
  booktitle={2015 IEEE International Conference on Communication Workshop (ICCW)}, 
  title={Wideband communication with high-dimensional arrays: New results and transceiver architectures}, 
  year={2015},
  volume={},
  number={},
  pages={1042-1047},
  doi={10.1109/ICCW.2015.7247314}}

@ARTICLE{ShenAnalog,
  author={Shen, Wenqian and Bu, Xiangyuan and Gao, Xinyu and Xing, Chengwen and Hanzo, Lajos},
  journal={IEEE Transactions on Signal Processing}, 
  title={Beamspace Precoding and Beam Selection for Wideband Millimeter-Wave {MIMO} Relying on Lens Antenna Arrays}, 
  year={2019},
  volume={67},
  number={24},
  pages={6301-6313},
  doi={10.1109/TSP.2019.2953595}}

@INPROCEEDINGS{localLMMSEAbdelghany,
  author={Abdelghany, Mohammed and Madhow, Upamanyu and Tölli, Antti},
  booktitle={2019 IEEE 20th International Workshop on Signal Processing Advances in Wireless Communications (SPAWC)}, 
  title={Beamspace Local {LMMSE}: An Efficient Digital Backend for mmWave Massive {MIMO}}, 
  year={2019},
  volume={},
  number={},
  pages={1-5},
  doi={10.1109/SPAWC.2019.8815585}}

@ARTICLE{layeredbeliefpropbeamspace,
  author={Takahashi, Takumi and Tölli, Antti and Ibi, Shinsuke and Sampei, Seiichi},
  journal={IEEE Transactions on Wireless Communications}, 
  title={Low-Complexity Large {MIMO} Detection via Layered Belief Propagation in Beam Domain}, 
  year={2022},
  volume={21},
  number={1},
  pages={234-249},
  doi={10.1109/TWC.2021.3094970}}

@article{charbonnier2020calibration,
  title={Calibration of ray-tracing with diffuse scattering against 28-{GH}z directional urban channel measurements},
  author={Charbonnier, Romain and Lai, Chiehping and Tenoux, Thierry and Caudill, Derek and Gougeon, Gr{\'e}gory and Senic, Jelena and Gentile, Camillo and Corre, Yoann and Chuang, Jack and Golmie, Nada},
  journal={IEEE Transactions on Vehicular Technology},
  volume={69},
  number={12},
  pages={14264--14276},
  year={2020},
  publisher={IEEE}
}

@book{Bhatia2007,
  title     = {Positive Definite Matrices},
  author    = {Bhatia, Rajendra},
  year      = {2007},
  publisher = {Princeton University Press},
  address   = {Princeton, New Jersey},
  series    = {Princeton Series in Applied Mathematics},
  isbn      = {978-0-691-12918-1}
}

@article{hansen1982jensen,
  title={Jensen's inequality for operators and L{\"o}wner's theorem},
  author={Hansen, Frank and Kj{\ae}rg{\aa}rd Pedersen, Gert},
  journal={Mathematische Annalen},
  volume={258},
  number={3},
  pages={229--241},
  year={1982},
  publisher={Springer}
}

\end{document}